\documentclass[10pt,journal,compsoc]{IEEEtran}

\usepackage[nocompress]{cite}
\usepackage{url}
\usepackage[pdftex]{graphicx}
\usepackage[cmex10]{amsmath}
\usepackage{algorithm}
\usepackage{algorithmic}
\usepackage{array}
\usepackage[caption=false,font=footnotesize,labelfont=sf,textfont=sf]{subfig}
\usepackage{fixltx2e}
\usepackage{stfloats}
\usepackage{multirow}
\usepackage{color}

\newtheorem{theorem}{Theorem}

\newtheorem{definition}{Definition}

\newtheorem{proof}{Proof}

\newcommand{\bx}{\boldsymbol{x}}

\begin{document}

\title{Time and Location Aware Mobile Data Pricing}


\author{Qian~Ma,~\IEEEmembership{Student Member,~IEEE,}
        Ya-Feng~Liu,~\IEEEmembership{Member,~IEEE,}
        and~Jianwei~Huang,~\IEEEmembership{Senior~Member,~IEEE}
\IEEEcompsocitemizethanks{
\IEEEcompsocthanksitem Q. Ma and J. Huang (corresponding author) are with the Network Communications and Economics Lab, Department of Information Engineering, the Chinese University of Hong Kong.\protect\\
E-mail: mq012@ie.cuhk.edu.hk; jwhuang@ie.cuhk.edu.hk.
\IEEEcompsocthanksitem Y.-F. Liu is with the State Key Laboratory of Scientific and Engineering Computing, Institute of Computational Mathematics and Scientific/Engineering Computing, Academy of Mathematics and Systems Science, Chinese Academy of Sciences, Beijing, 100190, China.\protect\\
E-mail: yafliu@lsec.cc.ac.cn.
}
\thanks{Part of this work has been presented at the IEEE International Conference on Communications (ICC), Sydney, Australia, June 10-14, 2014\cite{ICC}. This work is supported by the General Research Funds (Project Number CUHK 412713) established under the University Grant Committee of the Hong Kong Special Administrative Region, China, and the Natural Science Foundation of China, Grants 11301516 and 11331012.}
}


\IEEEtitleabstractindextext{
\begin{abstract}
Mobile users' correlated mobility and data consumption patterns often lead to severe cellular network congestion in peak hours and hot spots. This paper presents an optimal design of time and location aware mobile data pricing, which incentivizes users to smooth traffic and reduce network congestion. We derive the optimal pricing scheme through analyzing a two-stage decision process, where the operator determines the time and location aware prices by minimizing his total cost in Stage I, and each mobile user schedules his mobile traffic by maximizing his payoff (i.e., utility minus payment) in Stage II. We formulate the two-stage decision problem as a bilevel optimization problem, and propose a derivative-free algorithm to solve the problem for any increasing concave user utility functions. We further develop low complexity algorithms for the commonly used logarithmic and linear utility functions. The optimal pricing scheme ensures a win-win situation for the operator and users. Simulations show that the operator can reduce the cost by up to $97.52\%$ in the logarithmic utility case and $98.70\%$ in the linear utility case, and users can increase their payoff by up to $79.69\%$ and $106.10\%$ for the two types of utilities, respectively, comparing with a time and location independent pricing benchmark. Our study suggests that the operator should provide price discounts at less crowded time slots and locations, and the discounts need to be significant when the operator's cost of provisioning excessive traffic is high or users' willingness to delay traffic is low.
\end{abstract}

}

\maketitle

\IEEEdisplaynontitleabstractindextext

\IEEEraisesectionheading{\section{Introduction}\label{sec:intro}}

\IEEEPARstart{C}{isco} has predicated that the global mobile data demand will grow at an anticipated annual growth rate of $61\%$ from 2013 to 2018 \cite{Cisco}, but the mobile cellular network capacity has only been growing with an annual rate of $29\%$  \cite{Bloomberg}. As a result, the total mobile data demand may surpass the total network capacity globally very soon \cite{CD}, which could lead to significant performance deterioration and customer satisfaction loss. In order to alleviate the tension between supply and demand, the cellular operators have been trying to increase the network capacity through adopting new communication technologies (such as shifting from 3G to 4G technologies) and obtaining more spectrum (such as utilizing the TV white space for cellular communications \cite{TVWhite}). Another equally promising approach is to use economics mechanisms such as pricing to shape the customer demand and fully utilize the existing network resources \cite{Dyaberi}.

One widely used pricing strategy for shaping cellular data traffic is the usage-based pricing. For example, AT\&T in the USA has adopted a tiered usage-based monthly pricing plan since 2010, with the current rate of charging \$20 per month for 300MB and \$30 per month for 3GB of data \cite{ATmon}. However, the current usage-based pricing scheme often computes mobile users' network usage once every month, and ignores the stochastic nature of traffic over time and location.

From the cellular operator's point of view, the aggregate mobile data traffic varies significantly with time and location, and there are easily identifiable peak hours and crowded locations (such as business hours at commercial buildings and night time in highly populated residential areas) \cite{peakhour} \cite{dataload}. In fact, a major cost for the cellular operator is to cope with the peak demands at certain time slots and locations; meanwhile, the network capacity is not fully utilized at other time slots and locations. If a pricing scheme is aware of such traffic stochastics and provides proper incentives for users to shift traffic away from these time slots and locations, it will lead to a win-win situation for both the operator and users.

Time and location aware pricing is not completely new in the industry. Some heuristic schemes of this type have already existed in practice, such as MTN's dynamic tariffing in Africa and Uninor's dynamic pricing in India, both of which are designed for pricing voice calls \cite{Invention}. The success of these existing practices, together with the exploding wireless data demand, motivates us to provide a rigorous holistic design of time and location aware pricing for wireless data traffic. Notice that cellular operators usually charge data traffic based on volume and charge voice calls based on call durations, hence the optimal pricing schemes for these two types of traffic will be very different.

The research results regarding time-aware (but location independent) pricing for mobile data traffic only emerged very recently. Reference \cite{TUBE} demonstrated the effectiveness of time-aware pricing in terms of encouraging users to shift traffic to later non-peak hours. Reference \cite{Proactive} illustrated the possibility to use time-aware pricing to encourage users to pre-download data before peak hours. However, neither of them exploited the spatial dynamics of the traffic.

The only result regarding location-aware data pricing is the experiments from AT\&T \cite{Regional}. This pricing scheme separates the whole network area into several regions, and optimizes the prices for each region independently. Reference \cite{Regional} demonstrated that a location-based pricing can reduce network congestion, but did not provide analysis regarding users' mobilities, the impact of pricing on users' payoffs, and the time dimension traffic variations.

%

The goal of this paper is to design a time and location aware pricing scheme to provide benefits to both the cellular operator and mobile users.  Our main results and contributions are summarized as follows.

\begin{itemize}
  \item \emph{Problem Formulation:} To the best of our knowledge, this is the first study regarding \emph{a holistic optimal design of mobile data pricing in both time and spatial domains}. We capture the interactions between the operator and users as a two-stage decision process, considering users' global and local mobility patterns in the spatial domain and users' delay preference in the time domain.
  \item  \emph{Optimal Algorithm Design:} We formulate the time and location aware pricing problem as a bilevel optimization problem. The solution of the problem depends on the choice of users' utility functions. We propose to use the derivative-free algorithm as a general approach to solve the bilevel optimization problem, with general increasing concave utility functions.
  \item \emph{Customized Low Complexity Algorithm Design:} We also propose easily implementable low complexity algorithms, for two commonly used  utility functions. In particular, we propose the nonmonotone spectral projected gradient algorithm for the logarithmic utility case, and an algorithm combining the penalty idea and the block coordinate descent strategy for the linear utility case.
  \item \emph{Significant Performance Improvement:} Simulations show that both the cellular operator and users benefit from the time and location aware pricing scheme. In the logarithmic utility case, the cellular operator reduces the extra cost for provisioning the peak traffic by $97.52\%$, and users increase their payoffs by $79.69\%$, comparing with a time and location independent pricing benchmark. In the linear utility case, the cellular operator reduces the cost by up to $98.70\%$, and users increase their payoffs by up to $106.10\%$.
  \item \emph{Industry Insights:} Simulation results show that the operator will generally provide price discounts at less crowded time slots and locations, and the discounts need to be significant when the operator's cost of provisioning excessive traffic is high or the users' willingness to delay traffic is low.
\end{itemize}


The rest of this paper is organized as follows. We introduce the two-stage decision model in Section \ref{sec:model}. In Section \ref{sec:solu}, we present several algorithms for solving both the general problem and the special cases with logarithmic and linear utility functions. We verify the effectiveness of the proposed pricing scheme and analyze the impact of system parameters in Section \ref{sec:simu}. We finally conclude in Section \ref{sec:conc}.

\section{System Model}\label{sec:model}

We consider a cellular mobile network, where the cellular operator determines prices and mobile users decide their mobile data consumptions based on the prices. We assume that mobile users are price-takers, who do not anticipate the impact of their demands on the operator's prices. Such a price-taking behavior is reasonable, as the number of subscribers is usually large for a single operator, and the impact of a single user on the entire network is negligible.

\begin{figure}
  \vspace{-5mm}
  \centering
   \includegraphics[width=0.4\textwidth]{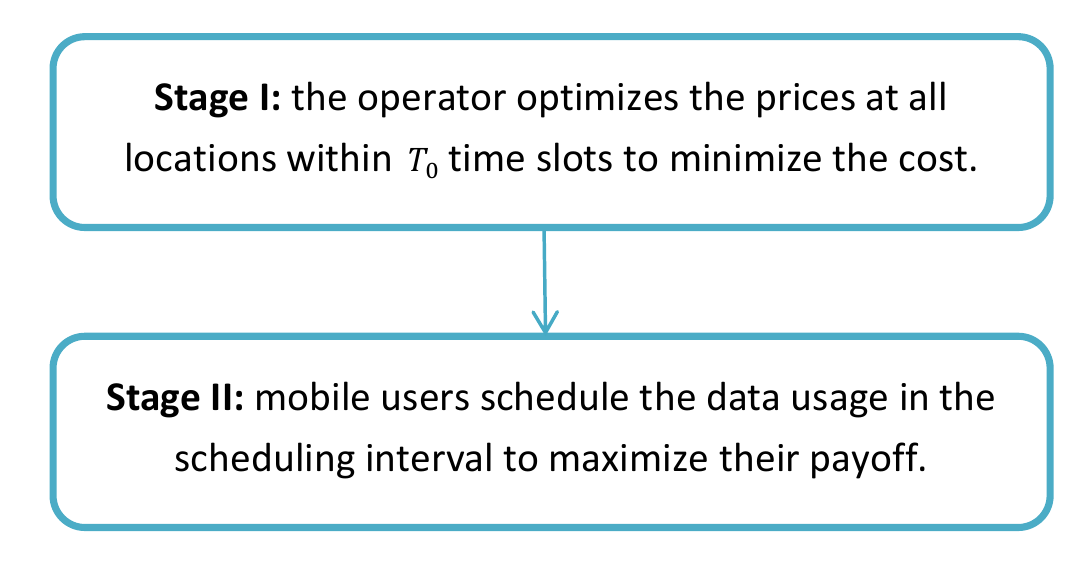}
   \vspace{-3mm}
  \caption{Two-stage decision process.}\label{fig:ArcFig}
  \vspace{-2mm}
\end{figure}

We capture the above sequential interactions between the operator and users as a two-stage decision process. During Stage I, the operator announces the prices for different time slots (e.g., different hours) and different locations (corresponding to the coverage areas of different base stations). In Stage II, each user decides his mobile data usage over time, based on the prices and his own mobility. Figure \ref{fig:ArcFig} shows the two-stage decision process. Notice that the existing commercial time and location independent usage based pricing scheme (such as the one used by AT\&T) is a special case of the more general model in this paper.

\begin{figure}
  \vspace{-2mm}
  \centering
   \includegraphics[width=0.47\textwidth]{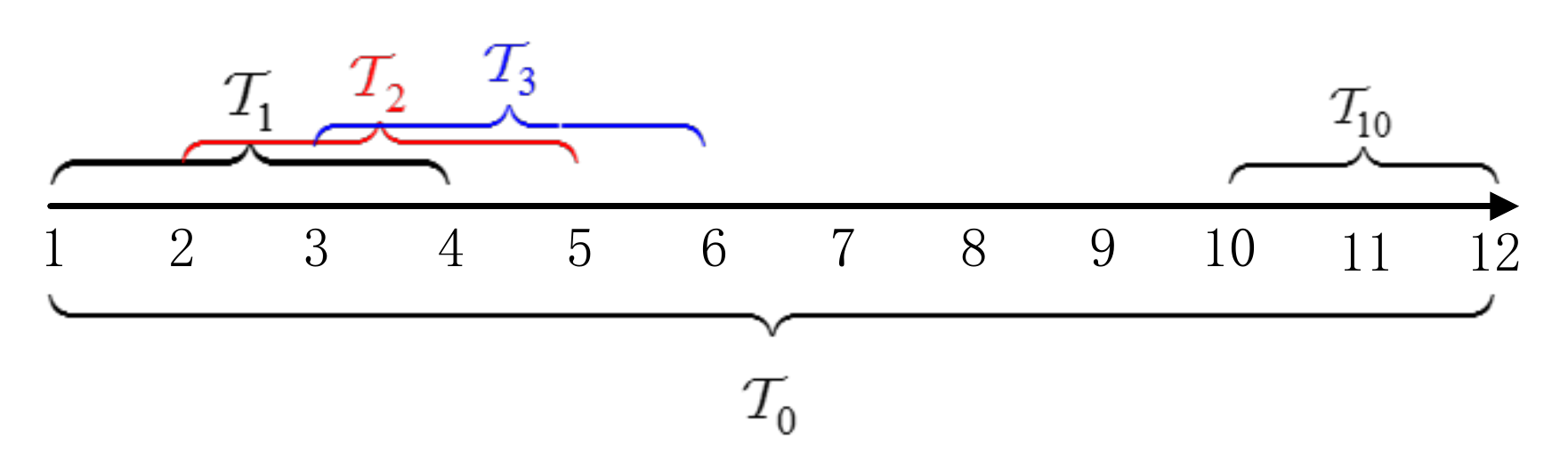}
   \vspace{-3mm}
  \caption{An example with $T_0=12$ and $T=4$. A user active at time slot $3$ can schedule his data at time slots $3,4,5,$ and $6$ in set $\mathcal{T}_3$. A user active at time slot $10$ can only schedule his data at time slots $10,11,$ and $12$ in set $\mathcal{T}_{10}$.}\label{fig:T0T}
  \vspace{-4mm}
\end{figure}


\emph{Time Domain Modeling:} In the time domain, the operator makes the pricing decisions in Stage I, for an entire period of $\mathcal{T}_0=\{1,2,\ldots ,T_0\}$ time slots. In Stage II, when a user becomes active in time $t\in\mathcal{T}_0$\footnote{For simplicity, we focus on the downlink transmission from the operator's base stations to the users in this paper. The uplink transmission can be analyzed similarly with a more detailed discussions about the interference management issues.}, he can schedule the mobile data consumption during one or more of the following time slots: $\mathcal{T}_t=\{t, t+1, \ldots , T_t\}$, where $T_t=\min\{t+T-1, T_0\}$. We call $T$ the scheduling interval\footnote{With a slight modification of the model, we can allow scheduling interval to be both user and time dependent. This will not change the major insights of our analysis.}, which is usually smaller than the operator's pricing span $T_0$. Figure \ref{fig:T0T} shows an example with $T_0=12$ and $T=4$.


\emph{Spatial Domain Modeling:} In the spatial domain, we consider users' mobility patterns, which capture their daily movement habits \cite{Traffic}. The cellular operator is able to construct aggregate \emph{mobility profiles} for the entire user population based on historical measurements \cite{M2}. There are two types of mobility profiles: the global (long-term) one which captures users' mobility at a larger time scale (say over a day or a week, which corresponds to $T_0$ time slots in this paper), and the local (short-term) one which captures users' mobility at a smaller time scale (say over several consecutive time slots, which corresponds to $T$ time slots in this paper).

We denote the set of locations in the network as $\mathcal{L} = \{1,2, \ldots , L\}$. The operator constructs the global mobility profile for all users:
\vspace{-2mm}
\begin{equation}\label{glo}
\boldsymbol{\alpha} = \{ \alpha(t,l):\alpha(t,l)\geq 0,\,\displaystyle\sum_{l=1}^L \alpha(t,l) =1,\,t\in \mathcal{T}_0,\,l \in \mathcal{L} \},
\end{equation}
where $\alpha(t,l)$ represents the probability of users appearing at location $l$ at time slot $t$ from a long-term point of view. A type $a$ user can make more precise predications for his mobility across time slots and locations through the local mobility profile:
\vspace{-3mm}
\begin{align}
\vspace{-4mm}
& \boldsymbol{\beta}_a = \Big\{ \beta_a(t',l'|t,l):\beta_a(t',l'|t,l)\geq 0,\,\sum_{l'=1}^L \beta_a(t',l'|t,l)=1, \notag\\
& \quad\quad\quad t\in\mathcal{T}_0,\,l\in\mathcal{L},\,t' \in \mathcal{T}_t\setminus\{t\},\,l' \in \mathcal{L} \Big\} .\label{loc}
\end{align}
Here  $\beta_a(t',l'|t,l)$ represents the probability of a type $a$ user appearing at location $l'$ at time slot $t'$, given that he has appeared at location $l$ at time slot $t$. We will explain the user type in more details in Section \ref{sec:user}. 
\emph{In this paper, we assume that users will follow their daily routines and will not change their mobility patterns by responding to the prices.}

Reference \cite{M2} provided more discussions regarding how the operator can construct the mobility profiles by learning users' movement history. In this paper, we assume that $\boldsymbol{\alpha}$ and $\boldsymbol{\beta}_a$ (for all $a$) are known system parameters\footnote{We also consider the case where the mobility profiles have errors. We use the idea of robust optimization to deal with the errors. Due to space limit, we put it in Appendix A.}.

Since the cellular operator will make price decisions based on users' responses to prices, we will analyze the two-stage decision process through backward induction.

\subsection{Users' Decision in Stage II}\label{sec:user}

In Stage II, a user needs to schedule his data usage to maximize his payoff (i.e., utility minus payment), given the prices announced by the operator in Stage I:
\begin{equation}\label{pricing}
\boldsymbol{p} = \left\{p(t,l):\,t \in \mathcal{T}_0,\,l \in \mathcal{L} \right\}.
\end{equation}
Here $\boldsymbol{p}$ is a $T_0 \times L$ matrix, and its $(t,l)$-th entry $p(t,l)$ denotes the price per unit of data traffic at time $t$ and location $l$.

A user's utility depends on several factors, including a \emph{utility function} that characterizes the user's satisfaction level of consuming certain amount of data traffic in a single time slot, the \emph{delay tolerance parameter} $\delta \in [0,1]$ that captures the user's willingness to wait, and the user's \emph{mobility pattern} which predicts his locations in the next $T$ time slots. We will divide the user population into a set $\mathcal{A}=\{1,2,\ldots,A\}$ of types\footnote{The operator can obtain the number of each type of users through historical information or long-term learning \cite{clustering}.}.
We define the user type as follows.
\begin{definition}[User Type]
Users with the same utility function, delay tolerance parameter, and mobility profile belong to the same user type.
\end{definition}

\subsubsection{Initial and Scheduled Demands}

Consider a type $a$ user who becomes active at time $t$ and location $l$ with a demand of $x_a^{ini}(t,l)>0$ data traffic. The superscript $ini$ represents that this is the ``initial'' demand before scheduling. If the prices announced by the operator in Stage I are time and location \emph{independent}, then the user will consume $x_a^{ini}(t,l)$ amount of traffic immediately in time slot $t$, as delaying the consumption will not increase the utility or decrease the payment.

When prices are time and location dependent, the user may choose to schedule the traffic to later time slots (and hence at possibly different locations based on his mobility) to maximize his payoff. We denote the traffic that a type $a$ user shifts from time $t$ and location $l$ to time $t'$ and location $l'$ as $x_a(t',l'|t,l)$. The traffic shift decisions form a vector:
\begin{equation}\label{schedule}
\boldsymbol{x}_a(t,l)=\left\{x_a(t,l|t,l),x_a(t',l'|t,l),t' \in \mathcal{T}_t\setminus\{t\},l'\in \mathcal{L}\right\}.
\end{equation}
Recall that $\mathcal{T}_t=\{t,t+1, \ldots , T_t\}$. Basically, if the user decides to use the data at the current time slot $t$, then the location can only be $l$ (which is known). We denote this amount of data as $x_a(t,l|t,l)$. If the user chooses to use the data at one of the future time slots in $\mathcal{T}_t \setminus \{t\}$, then the possible locations are determined by the mobility pattern.

Furthermore, we assume that a user's total demand does not change through scheduling (in the expected sense), i.e.,
\begin{equation}\label{initialusage}
x_a^{ini}(t,l)=x_a(t,l|t,l)+\sum_{t'=t+1}^{T_t} \sum_{l'=1}^L \beta_a(t',l'|t,l) x_a(t',l'|t,l).
\end{equation}
Here $\beta_a(t',l'|t,l)$ is the local mobility profile defined in \eqref{loc}.

\subsubsection{Utility, Payment, and Payoff Maximization}

Next we characterize how the user will calculate the utility, payment, and payoff based on the scheduled traffic.

We denote a type $a$ user's utility function as $u_a(\cdot)$.
Due to the principle of diminishing marginal returns \cite{logutility, logutility2}, we assume that the utility function is increasing and concave.
For mathematically simplicity, we further assume that $u_a(\cdot)$ is smooth  (continuously differentiable).

Under the ``initial'' data traffic consumption (before scheduling), a type $a$ user's utility by consuming $x_a^{ini}(t,l)$ data traffic is $u_a(x_a^{ini}(t,l))$.

Under the new consumption profile $\boldsymbol{x}_a(t,l)$ (after scheduling), a type $a$ user's new perceived utility is calculated by the \emph{Discounted Utility Model} (DUM)\footnote{As an example, 
we consider a user who watches a movie, which corresponds to the consumption of $x_m$ amount of data.
Under the ``initial'' traffic consumption, he obtains a utility $u(x_m)$.
However, if he schedules to watch half of the movie immediately and the other half of the movie one hour later, his perceived utility is $u(0.5x_m)+\delta u(0.5x_m)$ for the new traffic consumption profile.
Here $\delta$ is the exponentially discounting parameter.} in behavioral economics \cite{BehavioralEconomicsDUM}.
The discounted utility accounts for the future discounted value of a good in its present value, by \emph{exponentially} discounting the value according to the delay \cite{WikiDUM}. 
The DUM is commonly used in modeling users' intertemporal choice, as it captures users' psychological factors in time preference by a single discount rate.
Hence with the scheduled traffic, the user's utility for the consumption profile $\boldsymbol{x}_a(t,l)$ can be written as \eqref{EXPuti} (on the top of next page).

\begin{figure*}
\begin{align}
& U(\boldsymbol{x}_a(t,l)) = u_a(x_a(t,l|t,l)) \displaystyle+\sum_{t'=t+1}^{T_t}\sum_{l'=1}^L \beta_a(t',l'|t,l) \delta_a^{t'-t}u_a(x_a(t',l'|t,l)).\label{EXPuti}
\end{align}
\hrulefill \vspace*{4pt}
\vspace{-3mm}
\end{figure*}

\begin{figure*}[!t]
\vspace{-2mm}
\begin{align}
& P(\boldsymbol{x}_a(t,l)) = p(t,l)x_a(t,l|t,l) \displaystyle+\sum_{t'=t+1}^{T_t}\sum_{l'=1}^L \beta_a(t',l'|t,l) p(t',l')x_a(t',l'|t,l).\label{EXPpay}
\end{align}
\hrulefill \vspace*{4pt}
\end{figure*}

The utility $U(\boldsymbol{x}_a(t,l))$ captures the decrease of utility due to delay through the delay tolerance parameter $\delta_a \in[0,1]$.
Users of different types may have different delay tolerances. Users who are less patient will have a smaller delay tolerance parameter $\delta_a$, and are less willing to delay his traffic in exchange for a smaller payment.

For practical methods of estimating users' delay tolerance and utility, we refer interested readers to \cite{TUBE} and \cite{AMUSE}. In this paper, we assume that these parameters have been estimated accurately and  are known.

A user's (expected) usage-based payment with the scheduled traffic is calculated in \eqref{EXPpay}.

The user's objective is to maximize his payoff (i.e., utility minus payment) by choosing the best traffic scheduling decision. Mathematically, a type $a$ user who has an initial demand of $x_a^{ini}(t,l)$ at time $t$ and location $l$ needs to solve the following traffic scheduling problem:
\begin{align}
& \mbox{\textbf{Problem 1: User's Traffic Scheduling Problem}} \notag\\
& \displaystyle \mbox{max} ~~ U(\boldsymbol{x}_a(t,l)) - P(\boldsymbol{x}_a(t,l)) \notag\\
& \mbox{s.t.} ~~~~ \eqref{initialusage}~ \text{and}~ \boldsymbol{x}_a(t,l) \geq \boldsymbol{0} \label{con6}\\
& \mbox{var:} ~~~ \boldsymbol{x}_a(t,l) \mbox{ defined in \eqref{schedule}}. \notag
\end{align}
The constraint $\boldsymbol{x}_a(t,l) \geq \boldsymbol{0}$ in \eqref{con6} requires the scheduled data traffic vector $\boldsymbol{x}_a(t,l)$ to be component-wise nonnegative.

\begin{figure*}\label{KKT}
\begin{align}
& p(t,l)-u_a'(x_a(t,l|t,l))+\lambda_a(t,l)\geq0,\,\boldsymbol{x}_a(t,l)\geq \boldsymbol{0}, \label{kkt1}\\
& \beta_a(t',l'|t,l) \left[ p(t',l')-\delta_a^{t'-t}u_a'(x_a(t',l'|t,l))+ \lambda_a(t,l)\right] \geq 0,~t' \in \mathcal{T}_t\setminus\{t\},\,l' \in \mathcal{L}, \label{kkt2}\\
& \displaystyle x_a^{ini}(t,l)=x_a(t,l|t,l)+ \sum_{t'=t+1}^{T_t} \sum_{l'=1}^L \beta_a(t',l'|t,l) x_a(t',l'|t,l), \label{kkt4}\\
& x_a(t,l|t,l) \left[p(t,l)-u_a'(x_a(t,l|t,l))+\lambda_a(t,l)\right]=0, \label{kkt5}\\
& x_a(t',l'|t,l) \beta_a(t',l'|t,l) \left[p(t',l')-\delta_a^{t'-t}u_a'(x_a(t',l'|t,l)) + \lambda_a(t,l) \right]=0,~t' \in \mathcal{T}_t\setminus\{t\},\,l' \in \mathcal{L}. \label{kkt6}
\end{align}
\hrulefill \vspace*{4pt}
\vspace{-2mm}
\end{figure*}

Since the utility function $u_a(\cdot)$ is smooth and concave, Problem 1 is a smooth convex optimization problem. Therefore, the KKT conditions of Problem 1 shown in \eqref{kkt1}--\eqref{kkt6} (on the top of next page) are sufficient and necessary for its global optimality. Here $\lambda_a(t,l)$ is the Lagrangian multiplier associated with the equality constraint \eqref{initialusage} in Problem 1. We can see from the KKT conditions that if $\beta_a(t',l'|t,l)=0$, then $x_a(t',l'|t,l)=0$ is a solution. Intuitively, if the mobility pattern suggests that the user will never go to a position $l'$ at time slot $t'$, e.g., $\beta_a(t',l'|t,l)=0$, then naturally the user will set $x_a(t',l'|t,l)=0.$

Notice that if the utility function is strictly concave (e.g., the logarithmic utility function), then the optimal solution of Problem 1 is unique; while if the utility function is not strictly concave (e.g., the linear utility function), then the optimal solution of Problem 1 might not be unique, which implies that a user may have more than one optimal scheduling decision. To overcome this technical difficulty, we assume that the operator can guide the user to choose one particular solution that the operator prefers (if Problem 1 has multiple optimal solutions)\footnote{One way to achieve this is that the operator provides recommendation to the user through a mobile app. 
The operator can compute the optimal scheduling solution for the user, and send the best recommendation to the user.
 The TUBE mobile app designed in \cite{TUBE} can be used to achieve the above functionalities.}. This does not affect the user's maximum achievable payoff, but will make discussions later on considerably cleaner.

\subsection{The Cellular Operator's Decision in Stage I}\label{sec:operator}

In Stage I, the operator needs to optimize the time and location dependent prices to minimize his cost, considering the impact on the users' scheduling decisions in Stage II.

\subsubsection{The Cellular Operator's Cost}

We will consider two types of cost for the network operator: the cost of provisioning demand exceeding capacity, and the price discounts offered to the users as incentives.

\emph{Cost of Previsioning Excessive Demand}: When the data traffic exceeds the network capacity at a particular time slot and location, the operator will incur a significant additional cost to accommodate the extra traffic. Such a cost can be in two forms: (i) some of the traffic may not be delivered immediately, hence the users will experience a degraded Quality-of-Service due to an excessive delay, which in turn may lead to user churn and reduce the operator's revenue in the long run; (ii) the operator may need to obtain additional network resources at an extra cost, such as offloading to WiFi networks belonging to a different operator, or temporally leasing spectrum from other cellular operators \cite{extracost}. When the total scheduled user demand (from all user types) at time slot $t$ and location $l$ is $x^{aft}(t,l)$ (calculated in \eqref{aftusage}, on the top of next page), the cost of satisfying additional demand exceeding a capacity $C$ is \cite{TUBE}:
\begin{equation*}
f(x^{aft}(t,l)) = \gamma \max\left\{x^{aft}(t,l)-C,0\right\}.
\end{equation*}
Here $\gamma$ is the cost for serving an additional unit of traffic beyond the capacity.

\begin{figure*}\label{aftx}
\vspace{-2mm}
\begin{align}
& x^{aft }(t,l) = \sum_{a=1}^A x_a^{aft }(t,l)=\sum_{a=1}^A \left[ x_a (t,l|t,l) +\sum_{t''=\max\left\{t-T+1,1\right\}}^{t-1}\sum_{l''=1}^L \beta_a(t,l|t'',l'') x_a(t,l|t'',l'')\right],~t\in \mathcal{T}_0,\,l \in \mathcal{L}.\label{aftusage}
\end{align}
\hrulefill \vspace*{4pt}
\end{figure*}

\emph{Cost due to Price Discount}: When the cellular operator incentivizes users to shift traffic to less crowded time slots and locations through offering a price discount, the operator also experiences a loss of revenue. This can be viewed as another type of cost. Let us consider a benchmark flat-rate usage-based pricing $p_0$, which is time and location independent following most operators' practice today. We assume that the cellular operator can only provide discounts, but cannot charge prices higher than the benchmark  (i.e., $p(t,l) \leq p_0$). Recall our time and location aware prices are given in \eqref{pricing}. Then the discount at time $t$ and location $l$ is $p_0-p(t,l) \geq 0$. The constraint of providing discounts ensures that the new pricing scheme can only reduce the cost of the users, and hence will be embraced by users and supported by regulators during actual implementation. Moreover, since not providing any discounts is a feasible choice, the operator will not experience a total cost higher than today's time and location independent pricing benchmark. Hence the ``discount-only'' pricing scheme leads to a win-win situation for both the operator and users.

When the price for time slot $t$ and location $l$ is $p(t,l)$, the loss of revenue of serving the users' scheduled traffic at time slot $t$ and location $l$ is:
\begin{equation}
\left(p_0-p(t,l)\right)x^{aft}(t,l) . \label{cost}
\end{equation}
Here $x^{aft}(t,l)$ is all users' usage in time slot $t$ at location $l$ after scheduling.

\subsubsection{The Cellular Operator's Price Optimization}

The cellular operator's goal is to minimize his expected total cost across all time slots, locations, and user types, considering the global mobility pattern $\boldsymbol{\alpha}$ defined in \eqref{glo}.

Let us denote the optimal solutions of Problem 1 as $\boldsymbol{x}_a^\ast(t,l)$ for $t \in \mathcal{T}_0,$ $l \in \mathcal{L}$, and $a\in\mathcal{A}$. Notice that $\boldsymbol{x}_a^\ast(t,l)$ depends on the price $\boldsymbol{p}$, that is, it is a function of $\boldsymbol{p}$. After scheduling, the \emph{optimal} total amount of usage at time slot $t$ and location $l$ of all types of users, denoted as $x^{aft \ast}(t,l)$, can be calculated by \eqref{aftusage}, given $\boldsymbol{x}_a^\ast(t,l)$.
\begin{align}
& \mbox{\textbf{Problem 2: The Operator's Price Optimization Problem}} \notag\\
& \displaystyle \mbox{min} ~~ \displaystyle
\sum_{t=1}^{T_0}\sum_{l=1}^L\alpha(t,l) \left[ f\left(x^{aft \ast}(t,l)\right) - p(t,l)x^{aft \ast}(t,l) \right] \notag\\
& \mbox{s.t.} 
~~~ 0  \leq  p(t,l)  \leq  p_0,~t \in \mathcal{T}_0,\,l \in \mathcal{L} \label{23}\\
& \mbox{var:} ~~ p(t,l),~t \in \mathcal{T}_0,\,l \in \mathcal{L}. \notag
\end{align}

Some remarks on Problem 2 are as follows. First, we remove the term $p_0 \sum_{t,l}x^{aft \ast}(t,l)$ from the objective of Problem 2, compared with \eqref{cost}. This is because scheduling does not change the total traffic, i.e.,
\begin{equation*}
p_0 \sum_{t,l}x^{aft \ast}(t,l)=p_0 \sum_{t,l,a}x_a^{aft \ast}(t,l) = p_0 \sum_{t,l,a}x_a^{ini}(t,l).
\end{equation*}
Second, since $\boldsymbol{x}_a^\ast(t,l)$ is a function of the price $\boldsymbol{p}$, it follows that the price $\boldsymbol{p}$ is the only decision variable of Problem 2. Third, whether we can obtain the explicit expression of $\boldsymbol{x}_a^\ast(t,l)$ in terms of $\boldsymbol{p}$ depends on the utility function $u_a(\cdot)$. It turns out that if the utility functions are logarithmic and linear functions, we can obtain the closed-form expression of $\boldsymbol{x}_a^\ast(t,l)$ with respect to $\boldsymbol{p}.$ We will further discuss this in Sections \ref{sec:homo1} and \ref{sec:homo2}. For general concave utility functions, we can combine Problems 1 and 2, and reformulate it as an equivalent bilevel problem.

\subsection{Problem Reformulation}\label{sec:formulation}

The two-stage decision problems (Problem 1 and Problem 2) can be equivalently reformulated as a bilevel optimization problem. In a bilevel optimization problem \cite{bilevel}, a lower-level problem is embedded into an upper-level optimization problem. In this paper, the cellular operator's pricing problem (Problem 2) is the upper-level one, and the users' scheduling problem (Problem 1) is the lower-level one.

When the lower-level problem is convex, its optimal solution can be characterized by the necessary and sufficient KKT conditions, which can be embedded into the high-level problem and lead to the bilevel optimization formulation. By substituting the KKT conditions \eqref{kkt1}--\eqref{kkt6} into the operator's pricing Problem 2, we obtain the bilevel problem:
\begin{align}
& \mbox{\textbf{Problem 3: Bilevel Pricing and Scheduling Problem}} \notag\\
& \displaystyle \mbox{min}~~ \displaystyle
\sum_{t=1}^{T_0}\sum_{l=1}^L\alpha(t,l)\left[f\left(x^{aft}(t,l)\right)-p(t,l)x^{aft}(t,l)\right] \notag \\
& \mbox{s.t.}~~~\eqref{kkt1}-\eqref{kkt6},\eqref{aftusage},\eqref{23} \notag \\
& \mbox{var:}~p(t,l),\,\lambda_a(t,l),\,\boldsymbol{x}_a(t,l),\,x^{aft}(t,l),\,t \in \mathcal{T}_0,\,l \in \mathcal{L},\,a \in \mathcal{A}. \notag
\end{align}

Problem 3 is a nonconvex quadratic program. 

Key notations of our model are summarized in Table \ref{table-notation}.

\begin{table}[!ht]
\newcommand{\tabincell}[2]{\begin{tabular}{@{}#1@{}}#2\end{tabular}}
\caption{\textsc{Key Notations}} \label{table-notation}
\centering
\begin{tabular}[h]{|c|c|c|}
  \hline\hline
{Symbol} & {Physical Meaning} & {Eq.}
\\\hline
        {$\boldsymbol{\alpha}$}  &  {Global mobility profile}  & {\eqref{glo}}
\\\hline
        {$\boldsymbol{\beta}_a$}  &  {Local mobility profile of the type $a$ user}  & {\eqref{loc}} \\
\hline  {$\boldsymbol{p}$}  &   {Price announced by the operator} & {\eqref{pricing}} \\
\hline  {$x_a^{ini}(t,l)$}  & \tabincell{c}{Initial demand of the type $a$ user \\ at time slot $t$ and location $l$ before scheduling}  & {\eqref{initialusage}} \\
\hline  {$\boldsymbol{x}_a(t,l)$}  &  \tabincell{c}{Traffic shift decision vector of the type $a$ user \\ at time slot $t$ and location $l$} & {\eqref{schedule}} \\
\hline  {$x^{aft}(t,l)$}  &   \tabincell{c}{Total scheduled demand from all types of users\\at time slot $t$ and location $l$ after scheduling} & {\eqref{aftusage}} \\
\hline  {$\lambda_a(t,l)$}  &   {Lagrangian multiplier associated with \eqref{initialusage}} & {\eqref{kkt1}--\eqref{kkt6}} \\
\hline\hline
\end{tabular}
\end{table}

\section{Model Solution}\label{sec:solu}

In this section, we propose efficient algorithms for solving the bilevel pricing and scheduling problem (Problem 3). In Section \ref{sec:framework}, we first present a general solution approach which can be applied to solve the heterogenous case with any increasing concave utility functions. Then in Sections \ref{sec:homo1} and \ref{sec:homo2}, we propose customized low complexity algorithms for two special homogenous utility cases by judiciously exploiting the structures of the problem.

\subsection{A General Solution Approach}\label{sec:framework}

In this subsection, we propose a general solution approach for Problem 3, which can be applied to the heterogenous case with general increasing concave utility functions.

In Problem 3, the design variables are data usage $\boldsymbol{x}_a(t,l),x^{aft}(t,l)$, KKT multipliers $\boldsymbol{\lambda}_a$, and price $\boldsymbol{p}$ for all $t \in \mathcal{T}_0,\,l \in \mathcal{L},\,a \in \mathcal{A}.$ In fact, we know from \eqref{kkt1}--\eqref{kkt6},\eqref{aftusage} that $\boldsymbol{x}_a(t,l),x^{aft}(t,l)$ and $\boldsymbol{\lambda}_a$ are all functions of $\boldsymbol{p}$. Hence, Problem 3 can be seen as an optimization problem with regard to $\boldsymbol{p}$ by eliminating the variables $\boldsymbol{x}_a(t,l),x^{aft}(t,l)$ and $\boldsymbol{\lambda}_a.$ More specifically, Problem 3 is equivalent to the following Problem 4, with the variable being $\boldsymbol{p}$.
\begin{align}
& \mbox{\textbf{Problem 4}} \notag\\
& \displaystyle \mbox{min} ~~ H(\boldsymbol{p}) \notag \\
& \mbox{s.t.} ~~~\eqref{23} \notag \\
& \mbox{var:} ~~  p(t,l),~t \in \mathcal{T}_0,~l \in \mathcal{L}, \notag
\end{align}
where $H(\boldsymbol{p})$ is the optimal value of the following problem
\begin{align}
& \mbox{\textbf{Problem 5}} \notag\\
& \displaystyle \mbox{min} ~~ \displaystyle
\sum_{t=1}^{T_0}\sum_{l=1}^L\alpha(t,l)\left[f\left( x^{aft}(t,l)\right)-p(t,l)x^{aft}(t,l)\right] \notag \\
& \mbox{s.t.} ~~\eqref{kkt1}-\eqref{kkt6},\eqref{aftusage} \notag \\
& \mbox{var:} ~~\lambda_a(t,l),\,\boldsymbol{x}_a(t,l),x^{aft}(t,l),t \in \mathcal{T}_0,\,l \in \mathcal{L},\,a \in \mathcal{A}. \notag
\end{align}

Problem 4 is a box-constrained optimization problem. Next, we discuss the differentiability of $H(\boldsymbol{p})$ with respect to $\boldsymbol{p}$ and the calculation of $H(\boldsymbol{p})$ for a given particular $\boldsymbol{p},$ since both of them play important roles in solving Problem 4.

The objective function $H(\boldsymbol{p})$ of Problem 4 might not be differentiable for some choices of utility functions $u_a(\cdot)$.
In fact, when the utility function $u_a(\cdot)$ is linear, $H(\boldsymbol{p})$ in Problem 4 is discontinuous (and hence nondifferentiable) with respect to $\boldsymbol{p}.$ An illustrative example is given in Appendix B, which can be found in the supplemental material section of the manuscript center. Therefore, it is generally impossible to directly apply gradient-based methods to solve Problem 4.

Fortunately, computing $H(\boldsymbol{p})$ for any given $\boldsymbol{p}$, i.e., solving Problem 5 with a fixed $\boldsymbol{p}$, is relatively simple. Problem 5 is a convex optimization problem, since its objective function is convex, and the constraint set is convex and is composed of solutions of the users' traffic scheduling problem (Problem 1). More specifically, if the utility function $u_a(\cdot)$ is strictly concave (e.g., a logarithmic function), Problem 1 has a unique solution, so Problem 5 has a unique feasible point. Hence, solving Problem 5 is equivalent to solving convex Problem 1 and thus simple. If the utility function $u_a(\cdot)$ is linear, Problem 1 is a linear program, so the constraint of Problem 5 is a polyhedral set. Since the objective function of Problem 5 is piecewise linear with respect to $\boldsymbol{x}_a(t,l)$, Problem 5, after introducing some auxiliary variables, is a linear program. Hence, if the utility function $u_a(\cdot)$ is linear, solving Problem 5 is also simple.

The above analysis motivates us to use the derivative-free algorithm \cite{DeriFree} to solve Problem 4. The derivative-free algorithm, as a general solution approach, can solve Problem 4 with various types of increasing concave utility functions.

We propose to use the recently developed DYCORS (DYnamically COordinate search using Response Surface models) algorithm \cite{DYCORS}, which is one of the derivative-free algorithms, to solve Problem 4. The DYCORS algorithm is designed to solve the box-constrained large-scale optimization problem. The basic idea of the DYCORS algorithm is to build and maintain a surrogate model \cite{Surrogate} of the objective function at each iteration, and generate trial solutions by using a dynamic coordinate search strategy. The next iterate is selected from a set of random trial solutions obtained by perturbing only a subset of the coordinates of the current best solution, which is helpful in finding the global minimum. Moreover, the probability of perturbing a coordinate decreases as the algorithm reaches the computational budget.

In Algorithm \ref{algo:DYCORS}, we denote $n_0$ as the number of space-filling design points, $n$ as the number of previously evaluated points, $m$ as the number of trial points in each iteration, $\mathcal{A}_n=\{\boldsymbol{p}_1,\ldots ,\boldsymbol{p}_n\}$ as the set of previously evaluated points, and $s_n(\boldsymbol{p})$ as the response surface model built by using the points in $\mathcal{A}_n$. We denote $Nf_{max}$ as the maximum number of function evaluations allowed, and a strict decreasing function $\varphi(n)$ as the probability of perturbing a coordinate whose values are in $[0,1]$. Detailed discussions on the choice of these parameters can be found in \cite{DYCORS}.

\begin{algorithm}[h]
\caption{DYCORS Algorithm for Problem 4}
\label{algo:DYCORS}
\begin{algorithmic}[1]
\REQUIRE
problem inputs $p_0,T_0,L,A,\boldsymbol{\alpha},\gamma,C,\left\{\boldsymbol{\beta}_a,\delta_a\right\}_{a\in\cal A}$ and algorithm inputs $n_0,\mathcal{I}=\{\boldsymbol{p}_1,\ldots ,\boldsymbol{p}_{n_0}\},m,Nf_{max}$.
\ENSURE
$ \boldsymbol{p}^\ast. $
\STATE Evaluate $H(\boldsymbol{p})$ at the initial points $\mathcal{I}=\{\boldsymbol{p}_1,\ldots ,\boldsymbol{p}_{n_0}\}$.
\STATE Let $\boldsymbol{p}^{\ast}$ be the best point found so far. Set $n=n_0,\mathcal{A}_n=\mathcal{I}$.
\WHILE{$n < Nf_{max}$}
\STATE Fit/update a response surface model $s_n(\boldsymbol{p})$ using the data points in $\mathcal{B}_n=\{(\boldsymbol{p},H(\boldsymbol{p})):\boldsymbol{p}\in \mathcal{A}_n\}$.
\STATE Determine the probability $\varphi(n)$.
\STATE Generate trial points $\Omega_n=\{\boldsymbol{y}_{n,1},\ldots ,\boldsymbol{y}_{n,m}\}$ by:
\STATE $\mbox{ }$(1) Select the coordinates to perturb.
\STATE $\mbox{ }$(2) Randomly generate the trial points. 
\STATE $\mbox{ }$(3) Project the trial points onto the feasible set \eqref{23} (if necessary).
\STATE Select the next iterate $\boldsymbol{p}_{n+1}$ from $\Omega_n$ that minimizes $s_n(\boldsymbol{p})$.
\STATE Compute $H(\boldsymbol{p}_{n+1})$ by solving convex Problem 5.
\STATE If $H(\boldsymbol{p}_{n+1})< H(\boldsymbol{p}^{\ast})$, then $\boldsymbol{p}^{\ast}=\boldsymbol{p}_{n+1}$.
\STATE Set $\mathcal{A}_{n+1}=\mathcal{A}_n \cup \{\boldsymbol{p}_{n+1}\}$, and reset $n=n+1$.
\ENDWHILE
\end{algorithmic}
\end{algorithm}

\begin{theorem}
If the objective function of Problem 4 is continuous, then the sequences $\{\boldsymbol{p}_n\}$ generated by the DYCORS algorithm converge to the global minimum with probability one as $Nf_{\max} \rightarrow \infty$.
\end{theorem}

\begin{proof}
Detailed proof can be found in \cite{DYCORS}, Section 2.
\end{proof}

The DYCORS algorithm is powerful, in the sense that it can be used to solve a very general class of Problem 4 with various types of increasing concave utility functions. However, the algorithm often suffers slow convergence, as it does not exploit the special structures of the problem. Moreover, for the algorithm to converge to the global minimum, the objective of the corresponding problem is required to be continuous. This requirement, however, is not satisfied for Problem 4 with some choice of the utility function $u_a(\cdot)$ (e.g., the linear utility). In the following two subsections, we consider two special homogenous utility cases, and design tailored algorithms for Problem 4 by judiciously exploiting the corresponding problems' structures. According to the characteristics of users' wireless applications, the utility function $u_a(\cdot)$ can be either a linear function (such as file transfer \cite{utitwo}) or a strict concave function (elastic applications such as FTP and HTTP \cite{utitwo2}, \cite{utitwo3}). We first consider the homogeneous case where the utility functions are logarithmic, and then consider the homogeneous case where the utility functions are linear.

\subsection{Homogeneous Logarithmic Utility}\label{sec:homo1}

In this subsection, we study the homogeneous case where all users' utility functions are logarithmic. We assume that the utility function for a type $a$ user is
\vspace{-1mm}
\begin{equation*}
u_a(x) = k_a \log(1+x),
\vspace{-1mm}
\end{equation*}
where $k_a$ is a type specific parameter.

The user's traffic scheduling problem with the logarithmic utility function in Stage II is a strict convex problem. By solving its KKT conditions, we can obtain the explicit expression of $\boldsymbol{x}_a(t,l)$ in terms of the price $\boldsymbol{p}$ as follows:
\vspace{-1.5mm}
\begin{align}
x_a(t,l|t,l)&=\max \left\{\frac{k_a}{p(t,l)+\lambda_a (t,l)}-1,0\right\}, \label{Z1}\\
x_a(t',l'|t,l)&=\max \left\{\frac{k_a \delta_a^{t'-t}}{p(t',l')+\lambda_a (t,l)}-1,0\right\}, \label{Z2}
\end{align}
where $t'\in\mathcal{T}_t\setminus\{t\},\,l'\in\mathcal{L}$.
Here the KKT multiplier $\lambda_a (t,l)$ should be chosen such that \eqref{kkt4} is satisfied. By substituting \eqref{Z1} and \eqref{Z2} into \eqref{kkt4}, we see that the right hand side of \eqref{kkt4} is a monotonically decreasing function with respect to $\lambda_a (t,l)$. Hence we can use the binary search\footnote{The complexity of binary search is $\mathcal{O} (\log \frac{1}{\epsilon})$, where $\epsilon$ is the length of the final search interval. In our simulations, we set $\epsilon = 10^{-6}$. The impact of different choices of $\epsilon$ on the algorithm performance is presented in Appendix C.} to find the desired $\lambda_a^{\ast}(t,l)$ satisfying \eqref{kkt4}. The efficiency of the binary search heavily depends on the initial search interval containing $\lambda_a^{\ast}(t,l)$. Below we provide a lower bound and an upper bound of the desired $\lambda_a^{\ast} (t,l)$:
\vspace{-1mm}
\begin{equation*}
\frac{k_a}{x_a^{ini}(t,l)+1}-p(t,l) \leq \lambda_a^{\ast}(t,l) \leq k_a .
\vspace{-1mm}
\end{equation*}
Derivations of upper and lower bounds can be found in Appendix D.

For the logarithmic utility case, the objective function $H(\boldsymbol{p})$ in Problem 4 can be obtained by substituting the optimal $x_a^{\ast}(t,l|t,l)$ in \eqref{Z1} and $x_a^{\ast}(t',l'|t,l)$ in \eqref{Z2} into \eqref{aftusage}. Although in this case $H(\boldsymbol{p})$ in Problem 4 is continuous with respect to $\boldsymbol{p}$, it is still nondifferentiable due to the $\max$ operator in \eqref{Z1} and \eqref{Z2}. To take advantage of the explicit expressions of \eqref{Z1} and \eqref{Z2} and circumvent the difficulty of nondifferentiability of $H(\boldsymbol{p})$ with respect to $\boldsymbol{p},$ we propose a smoothing gradient-based method to solve Problem 4. In particular, we propose to smooth the objective $H(\boldsymbol{p})$ first, and then use the efficient spectral projected gradient method to solve the smoothed problem.

More specifically, we will approximate the $\max$ function $\theta(x)=\max\{x,0\}$ by the following smoothing function:
\begin{equation}
\tilde\theta(x;\mu)=\frac{1}{2}\left(x+\sqrt{x^2+\mu}\right), \label{apptheta}
\end{equation}
where the smoothing parameter $\mu$ is a sufficiently small positive number. It can be verified that
\begin{equation*}
0 \leq \tilde\theta(x;\mu)-\theta(x) \leq \frac{\sqrt{\mu}}{2}.
\end{equation*}
This implies that $\tilde\theta(x;\mu)$ uniformly converges to $\theta(x)$ as $\mu$ goes to zero\footnote{Uniform convergence implies that for any $\epsilon > 0$, there exists a $\bar \mu$ such that $|\tilde\theta(x;\mu) - \theta(x)|< \epsilon$, for all $\mu \leq \bar \mu$ and $x$.}.

\begin{figure*}
\vspace{-4mm}
\begin{equation}\label{Hmu}\tilde H(\boldsymbol{p};\mu)=\displaystyle \sum_{t=1}^{T_0}\sum_{l=1}^L\alpha(t,l)\left[\tilde f\left( \tilde x^{aft\ast}(t,l);\mu\right)-p(t,l) \tilde x^{aft\ast}(t,l)\right],\vspace{-2mm}\end{equation}
where
\vspace{-4mm}
\begin{align*}
& \tilde x^{aft \ast}(t,l) = \sum_{a=1}^A \tilde x_a^{aft \ast}(t,l) = \sum_{a=1}^A \left[ \tilde x_a^{ \ast}(t,l|t,l) +\sum_{t''=\max\left\{t-T+1,1\right\}}^{t-1}\sum_{l''=1}^L \beta_a(t,l|t'',l'') \tilde x_a^{\ast}(t,l|t'',l'')\right],~t\in \mathcal{T}_0,\,l \in \mathcal{L},\\
& \tilde x_a^{\ast}(t,l|t,l)=\tilde\theta\left(\frac{k_a}{p(t,l)+\tilde\lambda_a^{\ast} (t,l)}-1;\mu\right),~t\in \mathcal{T}_0,\,l \in \mathcal{L},\,a\in\mathcal{A}, \\
& \tilde x_a^{\ast}(t',l'|t,l)=\tilde\theta\left(\frac{k_a \delta_a^{t'-t}}{p(t',l')+\tilde\lambda_a^{\ast} (t,l)}-1;\mu\right),~t\in \mathcal{T}_0,\,l \in \mathcal{L},\,a\in\mathcal{A},\,t'\in\mathcal{T}_t\setminus\{t\},\,l'\in\mathcal{L}.
\end{align*}
\hrulefill \vspace*{4pt}
\vspace{-1mm}
\end{figure*}

\begin{figure*}
\vspace{-3mm}
\begin{equation}\label{tlambda}
x_a^{ini}(t,l)=\tilde x_a(t,l|t,l)+ \sum_{t'=t+1}^{T_t} \sum_{l'=1}^L \beta_a(t',l'|t,l) \tilde x_a(t',l'|t,l),~t\in \mathcal{T}_0,\,l \in \mathcal{L}.
\end{equation}
\hrulefill \vspace*{4pt}
\vspace{-2mm}
\end{figure*}

Using \eqref{apptheta}, we smooth the operator's cost function $f(\cdot)$ and the user's optimal scheduled traffic $\boldsymbol{x}_a(t,l)$, both containing the $\max$ operator. For notational simplicity, we denote the smoothed cost function and the scheduled traffic as $\tilde f(\cdot;\mu)$ and $\boldsymbol{\tilde x}_a(t,l)$, and the smoothed new usage as $\tilde{x}_a^{aft}(t,l)$ (which is a linear combination of $\boldsymbol{\tilde x}_a(t,l)$). We thus obtain a smoothed problem of Problem 4, which we denote as Problem 6:
\begin{align}
& \mbox{\textbf{Problem 6: Smoothed Problem}} \notag\\
& \displaystyle \mbox{min}~~ \tilde H(\boldsymbol{p};\mu) \notag \\
& \mbox{s.t.} ~~~\eqref{23} \notag \\
& \mbox{var:} ~~ p(t,l),~t \in \mathcal{T}_0,\,l \in \mathcal{L}. \notag
\end{align}
The objective function of Problem 6 is given in \eqref{Hmu} (on the next page), where $\tilde \lambda_a^{\ast}(t,l)$ should be chosen such that \eqref{tlambda} is satisfied. Note that $\tilde \lambda_a^{\ast}(t,l)$ can be computed by the binary search in the same fashion as $\lambda_a^{\ast}(t,l)$.

Problem 6 is a smooth box-constrained optimization problem. We propose to use the nonmonotone spectral projected gradient (SPG) algorithm (on the next page) \cite{PGM} to solve Problem 6. The pseudocode for the SPG algorithm is presented in Algorithm 2, where $\text{Proj}$ is the projection operator.

Three distinctive advantages of the SPG algorithm in the context of solving Problem 6 are as follows. First, the box constraint \eqref{23} is easy to project onto, and thus the SPG algorithm can be easily implemented to solve Problem 6. Second, since the SPG algorithm requires only the gradient information but not the high-order derivative information, it is suitable for solving large-scale optimization problems. Last but not least, the nonmonontone line search and the special choice of the stepsize make the SPG algorithm converge very fast and thus enjoy a quite good numerical performance. The nonmonontone line search (Line 6 in Algorithm 2) does not require the objective values monotonically decreasing at each iteration, which makes the trial points much easier to be accepted and is helpful in preventing the algorithm getting stuck at a local minimizer. The spectral stepsize ($\alpha_{k+1}$ in Line 16 of Algorithm 2) minimizes $\|\alpha \boldsymbol{s}_k-\boldsymbol{y}_k\|_2$, and thus provides a two-point approximation of the secant equation underlying Quasi-Newton methods \cite{taubook}, which often enjoy superlinear convergence rates.

\begin{figure*}
\vspace{-3mm}
\begin{equation} \label{gra}
\nabla \tilde H(\boldsymbol{p};\mu)=\displaystyle \sum_{t=1}^{T_0}\sum_{l=1}^L\alpha(t,l)\left[\nabla \tilde f\left(\tilde x^{aft\ast}(t,l);\mu\right)-\nabla p(t,l) \tilde x^{aft\ast}(t,l)- p(t,l) \nabla \tilde x^{aft\ast}(t,l) \right],\vspace{-3mm}
\end{equation}
where
\vspace{-2mm}
\begin{align}
& \nabla \tilde f\left(\tilde x^{aft\ast}(t,l);\mu\right)  = \frac{\gamma}{2}\left[ 1+\frac{\tilde x^{aft\ast}(t,l)-C}{\sqrt{\left(\tilde x^{aft\ast}(t,l)-C\right)^2+\mu}} \right] \nabla \tilde x^{aft\ast}(t,l),\notag\\
& \nabla \tilde x^{aft\ast}(t,l)=\sum_{a=1}^A \nabla \tilde x_a^{aft\ast}(t,l),\notag\\
& \nabla \tilde x_a^{aft \ast}(t,l)  = \nabla \tilde x_a^{\ast}(t,l|t,l) +\sum_{t''=\max\left\{t-T+1,1\right\}}^{t-1}\sum_{l''=1}^L \beta_a(t,l|t'',l'') \nabla \tilde x_a^{\ast}(t,l|t'',l''),\notag\\
& \nabla \tilde x_a^{\ast}(t,l|t,l) =\frac{-k_a\left(\nabla p(t,l)+\nabla \tilde\lambda_a^{\ast}(t,l)\right)}{2\left(p(t,l)+\tilde\lambda_a^{\ast}(t,l)\right)^2}\left[ 1+\displaystyle \frac{\frac{k_a}{p(t,l)+\tilde\lambda_a^{\ast}(t,l)}-1}{\sqrt{\left(\frac{k_a}{p(t,l)+\tilde\lambda_a^{\ast}(t,l)}-1\right)^2+\mu}} \right], \label{DZ1}\\
& \nabla \tilde x_a^{\ast}(t,l|t'',l'')=\frac{-k_a\delta_a^{t-t''}\left(\nabla p(t,l)+\nabla \tilde\lambda_a^{\ast}(t'',l'')\right)}{2\left(p(t,l)+\tilde\lambda_a^{\ast}(t'',l'')\right)^2}\left[ 1+\displaystyle\frac{\frac{k_a\delta_a^{t-t''}}{p(t,l)+\tilde\lambda_a^{\ast}(t'',l'')}-1}{\sqrt{\left(\frac{k_a\delta_a^{t-t''}}{p(t,l)+\tilde\lambda_a^{\ast}(t'',l'')}-1\right)^2+\mu}} \right].\label{DZ2}
\end{align}
\hrulefill \vspace*{4pt}
\vspace{-3mm}
\end{figure*}

\begin{theorem}\label{thrm2}
Any accumulation point\footnote{An accumulation point is a point which is the limit of a subsequence.} of the sequence generated by Algorithm 2 is a KKT point\footnote{A KKT point is a point which satisfies the KKT conditions of the optimization problem \cite{taubook}.} of Problem 6.
\end{theorem}
\begin{proof}
Detailed proof can be found in \cite{PGM}, Theorem 2.4 in Section 2 on Page 6. Note that Algorithm \ref{algo:SPG} in this paper is named SPG2 in \cite{PGM}.
\end{proof}

Now the only left question for applying the SPG algorithm (Algorithm \ref{algo:SPG}) to solve Problem 6 is the calculation of the gradient $\nabla \tilde H(\boldsymbol{p};\mu)$ of the smooth objective function $\tilde H(\boldsymbol{p};\mu).$ It turns out that $\nabla \tilde H(\boldsymbol{p};\mu)$ can be calculated by using the implicit function theorem (albeit the dependence of $\tilde \lambda_a^{\ast}(t,l)$ satisfying \eqref{tlambda} on $\boldsymbol{p}$ cannot be explicitly written). 
According to the composite rule of differentiation, we can calculate the gradient of the smoothed objective function in Problem 6 with respect to $\boldsymbol{p}$ as in \eqref{gra} (on the top of next page).
Notice that $\nabla p(t,l)=\mathbf{E}^{t,l}$ in \eqref{gra}, where $\mathbf{E}^{t,l}$ denotes the $T\times L$ matrix with all entries being zero except the $(t,l)$-th entry being one. Now, to compute $\nabla \tilde H(\boldsymbol{p};\mu)$, we only need to compute $\nabla \tilde \lambda_a^{\ast} (t,l)$ for $t\in \mathcal{T}$ and $l \in \mathcal{L}.$ Next, we apply the implicit function theorem to show the existence of $\nabla \tilde \lambda_a^{\ast} (t,l)$ and compute $\nabla \tilde \lambda_a^{\ast} (t,l).$

It is simple to check that the right hand side of \eqref{tlambda}, as a function of $\tilde \lambda_a(t,l)$, strictly decreases as $\tilde \lambda_a(t,l)$ increases. Hence, the derivative of the right hand side of \eqref{tlambda} with respect to $\tilde \lambda_a(t,l)$ is not equal to zero. By the implicit function theorem, we know $\nabla \tilde \lambda_a^{\ast} (t,l)$ exists. Then, by taking derivatives with respect to $\boldsymbol{p}$ on both sides of \eqref{tlambda}, we obtain
\begin{equation*}\label{7}
\nabla \tilde x_a^{\ast}(t,l|t,l) + \sum_{t'=t+1}^{T_t} \sum_{l'=1}^L \beta_a(t',l'|t,l) \nabla \tilde x_a^{\ast} (t',l'|t,l)=0.
\end{equation*}
Substituting \eqref{DZ1} and \eqref{DZ2} (on the next page) into the above equation, we can obtain $\nabla \tilde \lambda_a^{\ast} (t,l).$

\begin{algorithm}[h]
\caption{SPG Algorithm for Problem 6}
\label{algo:SPG}
\begin{algorithmic}[1]
\REQUIRE
problem inputs $p_0,T_0,L,A,\boldsymbol{\alpha},\gamma,C,\left\{\boldsymbol{\beta}_a,\delta_a\right\}_{a\in\cal A}$ and algorithm inputs $\alpha_0,\alpha_{\min},\alpha_{\max},M,\xi,\sigma_1,\sigma_2,\varepsilon,\mu$.
\ENSURE
$ \boldsymbol{p}_{\mu}. $
\STATE Let $\boldsymbol{p}_{\mu}=\boldsymbol{p}_0.$
\WHILE {$ ||\text{Proj}\left(\boldsymbol{p}_k-\nabla \tilde H(\boldsymbol{p}_k;\mu)\right)-\boldsymbol{p}_k|| > \varepsilon $}
\STATE Compute $\mathbf{d}_k=\text{Proj}\left(\boldsymbol{p}_k-\alpha_k \nabla \tilde H(\boldsymbol{p}_k;\mu)\right)-\boldsymbol{p}_k$. 
\STATE Set $\eta \leftarrow 1$.
\STATE Set $\boldsymbol{p}_+=\boldsymbol{p}_k + \eta \mathbf{d}_k$.
\WHILE {$\displaystyle \tilde H(\boldsymbol{p}_+;\mu) > \max_{0 \leq j \leq \min\{k,M-1 \}} \tilde H(\boldsymbol{p}_{k-j};\mu)$\\$~~~~~~~~~~~~~~~~~~~~+\xi \eta \big \langle \mathbf{d}_k,\nabla \tilde H(\boldsymbol{p}_k;\mu)\big \rangle$}
\STATE Find $\eta_{new} \in [\sigma_1 \eta, \sigma_2 \eta], $ set $\eta \leftarrow \eta_{new}$.
\STATE Set $\boldsymbol{p}_+=\boldsymbol{p}_k + \eta \mathbf{d}_k$.
\ENDWHILE
\STATE Let $\eta_k = \eta,~\boldsymbol{p}_{k+1}=\boldsymbol{p}_+,$\\$\boldsymbol{s}_k=\boldsymbol{p}_{k+1}-\boldsymbol{p}_k,~\boldsymbol{y}_k=\nabla \tilde H(\boldsymbol{p}_{k+1};\mu)-\nabla \tilde H(\boldsymbol{p}_k;\mu)$.
\STATE Compute $b_k=\langle \boldsymbol{s}_k,\boldsymbol{y}_k \rangle$.
\STATE If $\tilde H(\boldsymbol{p}_{k+1},\mu)< \tilde H(\boldsymbol{p}_{\mu})$, then $\boldsymbol{p}_{\mu}=\boldsymbol{p}_{k+1}$.
\IF {$b_k \leq 0$}
\STATE Set $\alpha_{k+1} = \alpha_{\max}$.
\ELSE
\STATE Compute $a_k = \langle \boldsymbol{s}_k,\boldsymbol{s}_k \rangle$ and\\
$ \alpha_{k+1} = \min\{ \alpha_{\max}, \max\{ \alpha_{\min}, a_k/b_k \} \}. $
\ENDIF
\ENDWHILE
\end{algorithmic}
\end{algorithm}

From the above analysis, as the parameter $\mu$ goes to zero, $\tilde H(\boldsymbol{p},\mu)$ \emph{uniformly} {converges to} $H(\boldsymbol{p})$. Moreover, the solution of Problem 6 also converges to the solution of Problem 4 \cite{smooth}. Therefore, when the parameter $\mu$ is very close to zero, the solution $ \boldsymbol{p}_{\mu}$ of Problem 6 returned by Algorithm 2 will be very close to the one of Problem 4.

Finally, when applying the SPG algorithm to solve Problem 6, we employ the continuation technique \cite{continuation}. That is, to obtain an approximate solution of Problem 4, we solve Problem 6 with a series of gradually decreasing values for $\mu$, instead of using a small fixed $\mu$. It turns out the continuation technique can reasonably improve the computational efficiency.

\subsection{Homogeneous Linear Utility}\label{sec:homo2}

In this subsection, we consider the homogeneous case with the linear utility function, i.e., $ u_a(x) = \rho_a x$, where $\rho_a$ is a type specific parameter. In this case, the users' traffic scheduling problem (Problem 1)  reduces to
\vspace{-1mm}
\begin{align}
& \displaystyle \mbox{max} ~~ \left(\rho_a-p(t,l)\right)x_a(t,l|t,l) \displaystyle  \label{linearproblem}\\
& ~~~ +\sum_{t'=t+1}^{T_t}\sum_{l'=1}^L  \left(\delta_a^{t'-t}\rho_a- p(t',l')\right)\beta_a(t',l'|t,l) x_a(t',l'|t,l) \notag\\
& \mbox{s.t.} ~~~ \eqref{initialusage}~ \text{and}~ \boldsymbol{x}_a(t,l) \geq \boldsymbol{0} \notag\\
& \mbox{var:} ~~ \boldsymbol{x}_a(t,l) \mbox{ defined in \eqref{schedule}}. \notag
\end{align}


The optimal solution of problem \eqref{linearproblem} is:
\begin{align}
& \{\boldsymbol{x}_a(t,l) \geq \boldsymbol{0}:  \boldsymbol{x}_a(t,l)~\text{satisfies}~\eqref{initialusage},~~~~~~~~~~~\notag \\
& ~~ \text{and}~x_a(t',l'|t,l)=0~\text{if}~\delta_a^{t'-t}\rho_a- p(t',l')<\upsilon(t,l)\}, \label{linearsolution}
\end{align}
where
\vspace{-2mm}
\begin{equation}\label{gamma}
\vspace{-1mm}
\upsilon(t,l)=\max_{(t',l')\in\{(t,l)\}\cup \{\mathcal{T}_t\setminus\{t\}\times \mathcal{L}\}}\left\{\delta_a^{t'-t}\rho_a- p(t',l')\right\}.
\end{equation}
If there is only one element in $\{(t,l)\}\cup \{\mathcal{T}_t\setminus\{t\}\times \mathcal{L}\}$ such that the maximum in \eqref{gamma} is achieved, then problem \eqref{linearproblem} has a unique solution; otherwise problem \eqref{linearproblem} has multiple solutions.

To overcome the computational difficulty of Problem 3 (with the linear utility function), we propose to penalize the complementarity constraints \eqref{kkt5} and \eqref{kkt6} to the objective function  with a parameter $\tau$. This transforms Problem 3 to Problem 7.
\vspace{-2mm}
\begin{align}
& \mbox{\textbf{Problem 7: Penalty-Based Problem}} \notag\\
& \displaystyle \min ~~ \displaystyle \sum_{t=1}^{T_0}\sum_{l=1}^L\alpha(t,l)\left[f\left(x^{aft}(t,l)\right)-p(t,l) x^{aft}(t,l)\right] \notag\\
& ~~~~~~+\displaystyle \tau \sum_{a=1}^A\sum_{t=1}^{T_0}\sum_{l=1}^{L} \left[\phi_a(t,l|t,l)+\sum_{t'=t+1}^{T_t}\sum_{l'=1}^L \phi_a(t',l'|t,l) \right] \notag\\
& \mbox{s.t.} ~~~ \eqref{kkt1}-\eqref{kkt4},\eqref{aftusage},\eqref{23} \notag\\
& \mbox{var:} ~  p(t,l),\lambda_a(t,l),\boldsymbol{x}_a(t,l),x^{aft}(t,l),t \in \mathcal{T}_0,l \in \mathcal{L}, a \in \mathcal{A}. \notag
\end{align}
Here $\phi_a(t,l|t,l)$ and $\phi_a(t',l'|t,l),$ corresponding to the complementarity constraints, are as follows:
\begin{gather}
\vspace{-2mm}
\begin{align}
& \phi_a(t,l|t,l)  = \left[p(t,l)-\rho_a+\lambda_a(t,l)\right]x_a(t,l|t,l), \notag\\
& \phi_a(t',l'|t,l)  = \left[p(t',l')-\delta_a^{t'-t}\rho_a +\lambda_a(t,l)\right]x_a(t',l'|t,l) \notag \\
& \quad \quad \quad \quad \quad \quad \cdot \beta_a(t',l'|t,l). \notag
\end{align}
\end{gather}

Problem 7 is equivalent to Problem 3 (with the linear utility function) as long as the penalty parameter $\tau$ is sufficiently large. This is because in Problem 7 we are trying to minimize the summation of two terms, one is the original objective function of the total cost in Problem 3, and the other is the penalized term of linear complementary constraints. Intuitively, when the penalty parameter is sufficient large, the latter one will dominate the former one, and we will minimize the latter one with higher priority and the former one with a lower priority. 
The above intuition is formally stated in the following theorem.
\begin{theorem}\label{thrm3}
There exists a $\tau_0>0$, such that Problem 7 with any $\tau \geq \tau_0$ and Problem 3 with the linear utility function share the same local minimizers and KKT points.
\end{theorem}
\begin{proof}
Detailed proof can be found in \cite{taubook}, Theorems 17.3 and 17.4 in Chapter 17.
\end{proof}

The threshold value $\tau_0$ in Theorem \ref{thrm3} is unknown in practice. However, we can choose it through a trial and error process. We can start with an initial estimation of $\tau$, solve Problem 7 (using Algorithm \ref{algo:BCD} discussed below) until convergence, and check whether the complementarity constraints \eqref{kkt5} and \eqref{kkt6} are satisfied at the solution. If yes, then we are done; otherwise we increase $\tau$ and then solve Problem 7 again.

Although Problem 7 is still nonconvex due to the nonconvex objective function, all of its constraints are convex, and the variables are decoupled in the constraints. This motivates us to use the block coordinate descent (BCD) algorithm to solve Problem 7 \cite{converge}. The key idea of the BCD algorithm is to partition variables in Problem 7 into two blocks: $\{\boldsymbol{p},\boldsymbol{\lambda}_a\}_{a \in \mathcal{A} }$ and $\{\boldsymbol{x}_a(t,l),x^{aft}(t,l)\}_{t \in \mathcal{T}_0,l \in \mathcal{L}, a \in \mathcal{A} }$. When we fix the variables in one block, Problem 7 becomes a linear programming problem of variables in the other block, and thus can be solved efficiently (to its optimality). Then we iteratively solve the variables in two blocks until the algorithm converges. Algorithm \ref{algo:BCD} provides the details of the BCD algorithm\footnote{The parameter $\varepsilon$ is the relative error, and $\varepsilon_0$ is the relative error tolerance which is a small positive number. In our simulations, we set $\varepsilon_0=10^{-6}$.}.

\begin{algorithm}[h]
\caption{BCD Algorithm for Problem 7}
\label{algo:BCD}
\begin{algorithmic}[1]
\REQUIRE
problem inputs $p_0,T_0,L,A,\boldsymbol{\alpha},\gamma,C,\left\{\boldsymbol{\beta}_a,\delta_a\right\}_{a\in\cal A}$ and algorithm inputs $\tau,\varepsilon_0$.
\ENSURE
$\boldsymbol{p}.$

\STATE Initialization: $\varepsilon = \infty$\;
\WHILE {$\varepsilon > \varepsilon_0$}
\STATE   Solve Problem 7 in terms of variables $\{\boldsymbol{p},\boldsymbol{\lambda}_a\}_{a \in \mathcal{A} }$, assuming other variables are fixed.
\STATE   Solve Problem 7 in terms of variables $\{\boldsymbol{x}_a(t,l),x^{aft}(t,l)\}_{t \in \mathcal{T}_0,l \in \mathcal{L},a \in \mathcal{A} }$, assuming other variables are fixed.
\STATE   $\varepsilon$ is the relative tolerance of the new and old $\boldsymbol{x}_a$.
\ENDWHILE
\end{algorithmic}
\end{algorithm}

The complexity of solving the linear program problem \cite{linearcomplexity} with respect to the variables $\{\boldsymbol{p},\boldsymbol{\lambda}_a\}_{a \in \mathcal{A} }$ (Line 3 in Algorithm \ref{algo:BCD}) is $O\left((T_0TL^2A)^{1.5}(T_0LA)^2\right),$ and the complexity of solving the linear program problem with respect to the variables $\{\boldsymbol{x}_a(t,l),x^{aft}(t,l)\}_{t \in \mathcal{T}_0,l \in \mathcal{L},a \in \mathcal{A} }$ (Line 4 in Algorithm \ref{algo:BCD}) is $O\left((T_0TL^2A)^{1.5}(T_0TL^2A)^2\right).$ The complexity of the entire BCD algorithm is still an open problem \cite{Complexity}. However, the BCD algorithm converges very fast in practice, and most of the objective improvement is achieved in the first few iterations.


The proposed BCD algorithm can be easily implemented, since we can use a mature linear programming solver to solve each step. The algorithm is guaranteed to converge to a KKT solution, which is in general the best we can do for the general NP-hard problems.

\begin{theorem}\label{thrm4}
The sequence generated by Algorithm 3 globally converges to a KKT point of Problem 7.
\end{theorem}
\begin{proof}
Proof can be found in \cite{converge}, Sections 4 and 5.
\end{proof}

\section{Simulation Results}\label{sec:simu}

\begin{table}[t]
\newcommand{\tabincell}[2]{\begin{tabular}{@{}#1@{}}#2\end{tabular}}
\caption{\textsc{Comparison between SPG and DYCORS (a smaller objective means a better performance)}} \label{table-com}
\centering
\begin{tabular}{|c|c|c|c|c|}
\hline
\hline
\multirow{2}{*}{\tabincell{c}{instances \\ $(T_0 \times L)$}} & \multicolumn{2}{c|}{SPG Algorithm} & \multicolumn{2}{c|}{DYCORS Algorithm} \\
\cline{2-5}
& obj. value & cpu time & obj. value & cpu time\\
\hline
\tabincell{c}{instance 1 \\ $(8 \times 3)$} & -476.4558 & 0.1404 & -476.1956 & 90.2466 \\
\hline
\tabincell{c}{instance 2 \\ $(8 \times 3)$} & -397.6181 & 0.0936 & -397.4685 & 90.3714 \\
\hline
\tabincell{c}{instance 3 \\ $(8 \times 3)$} & -527.4697 & 0.1092 & -527.2909 & 87.6570 \\
\hline
\tabincell{c}{instance 4 \\ $(10 \times 3)$} & -664.9654 & 0.1872 & -664.3879 & 126.6884 \\
\hline
\tabincell{c}{instance 5 \\ $(10 \times 3)$} & -447.4030 & 0.2028 & -447.1550 & 133.7241 \\
\hline
\tabincell{c}{instance 6 \\ $(10 \times 3)$} & -635.6959 & 0.1872 & -635.4643 & 111.1975 \\
\hline
\tabincell{c}{instance 7 \\ $(12 \times 3)$} & -776.1302 & 1.1544 & -775.4417 & 386.0401 \\
\hline
\tabincell{c}{instance 8 \\ $(12 \times 3)$} & -621.9934 & 1.0608 & -621.6543 & 392.7481 \\
\hline
\tabincell{c}{instance 9 \\ $(12 \times 3)$} & -544.1266 & 0.9984 & -543.8739 & 388.0246 \\
\hline
\tabincell{c}{instance 10 \\ $(24\times 3)$} & -1552.6 & 3.6192 & -1551.8 & 1845.0 \\
\hline
\tabincell{c}{instance 11 \\ $(24 \times 3)$} & -1330.4 & 3.2604 & -1329.9 & 2002.7 \\
\hline
\tabincell{c}{instance 12 \\ $(24 \times 3)$} & -1496.3 & 3.1512 & -1495.6 & 1870.9 \\
\hline\hline
\end{tabular}
\end{table}

In this section, we evaluate the performance of our proposed time and location aware pricing scheme for the two homogenous utility scenarios.\footnote{We also simulate a time-dependent pricing algorithm and compare it with our proposed time and location aware pricing scheme. Due to space limit, we put it in Appendix E.} For each scenario, we illustrate both the effectiveness of the proposed pricing scheme and the impact of various system parameters.

\subsection{Homogeneous Logarithmic Utility Scenario}

\subsubsection{The Effectiveness of Our Pricing Scheme}

\begin{figure*}
\vspace{-5mm}
\hspace{-5mm}
  \begin{minipage}[t]{0.33\linewidth}
  \centering
  \includegraphics[width=1\textwidth]{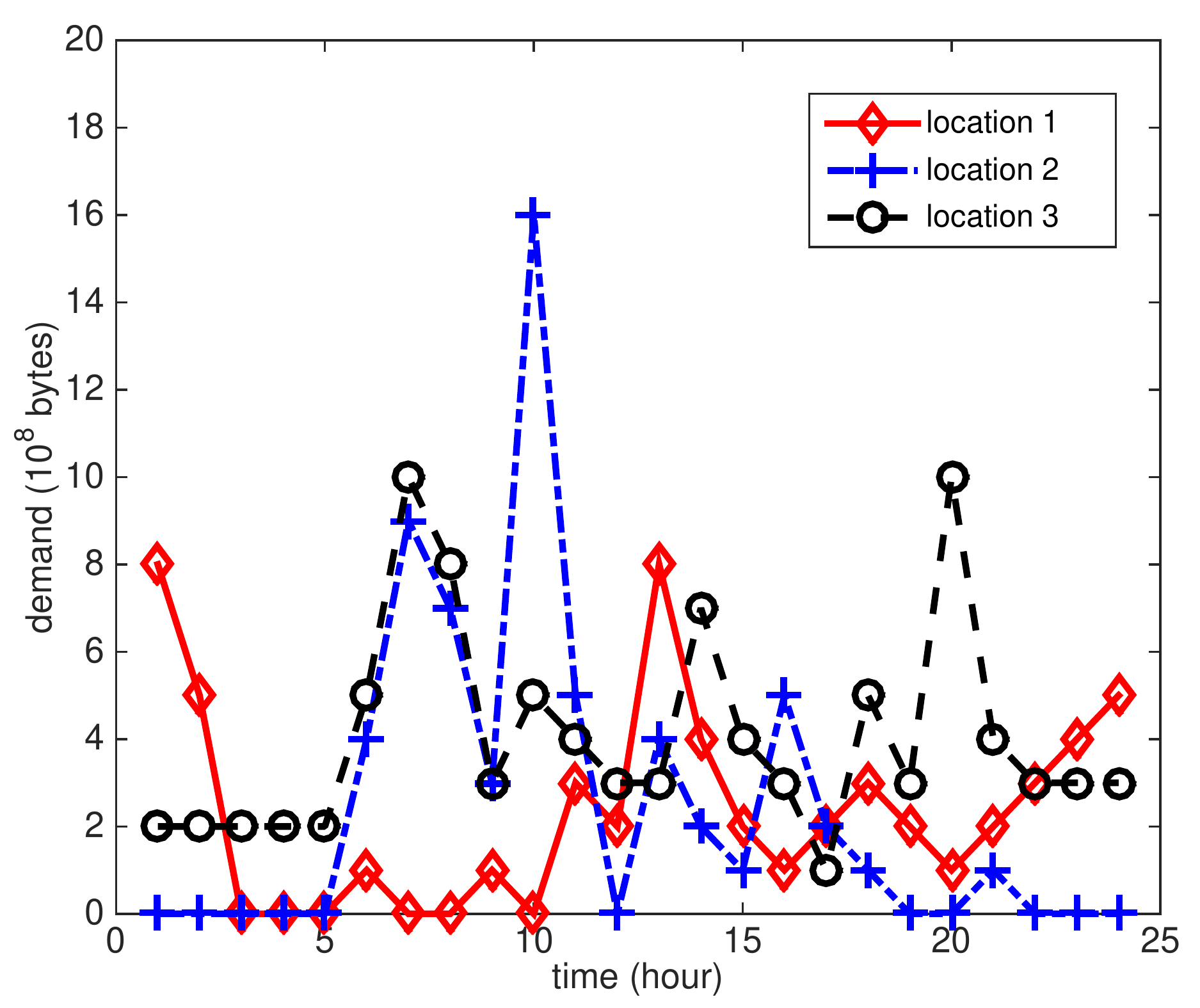}
  \caption{Initial traffic of 24 time slots.}\label{fig:ini2}
  \end{minipage}
    \begin{minipage}[t]{0.33\linewidth}
      \centering
      \includegraphics[width=1\textwidth]{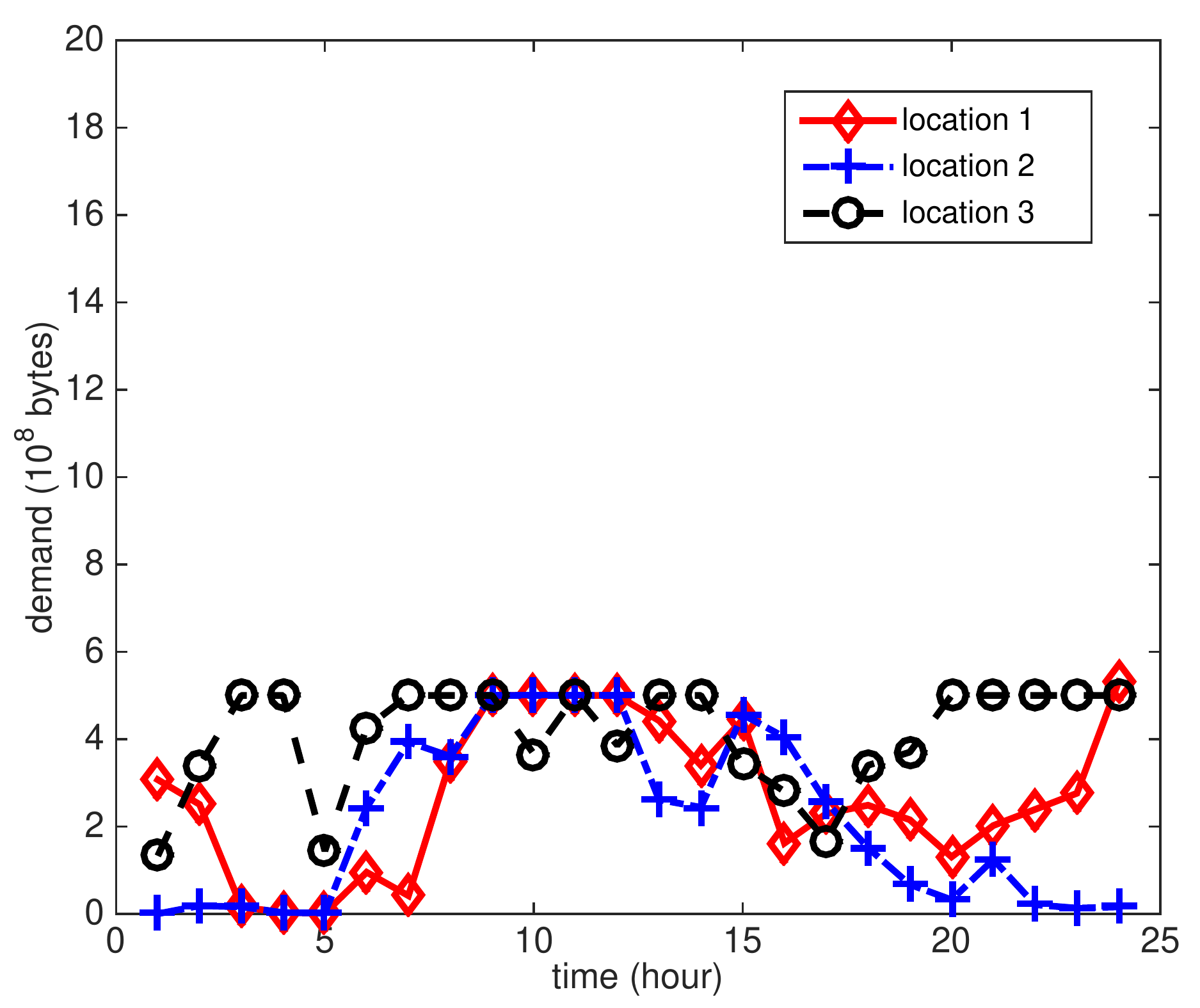}
      \caption{The shifted traffic under time and location dependent pricing.}\label{fig:logusa}
    \end{minipage}
  \begin{minipage}[t]{0.33\linewidth}
  \centering
  \includegraphics[width=1\textwidth]{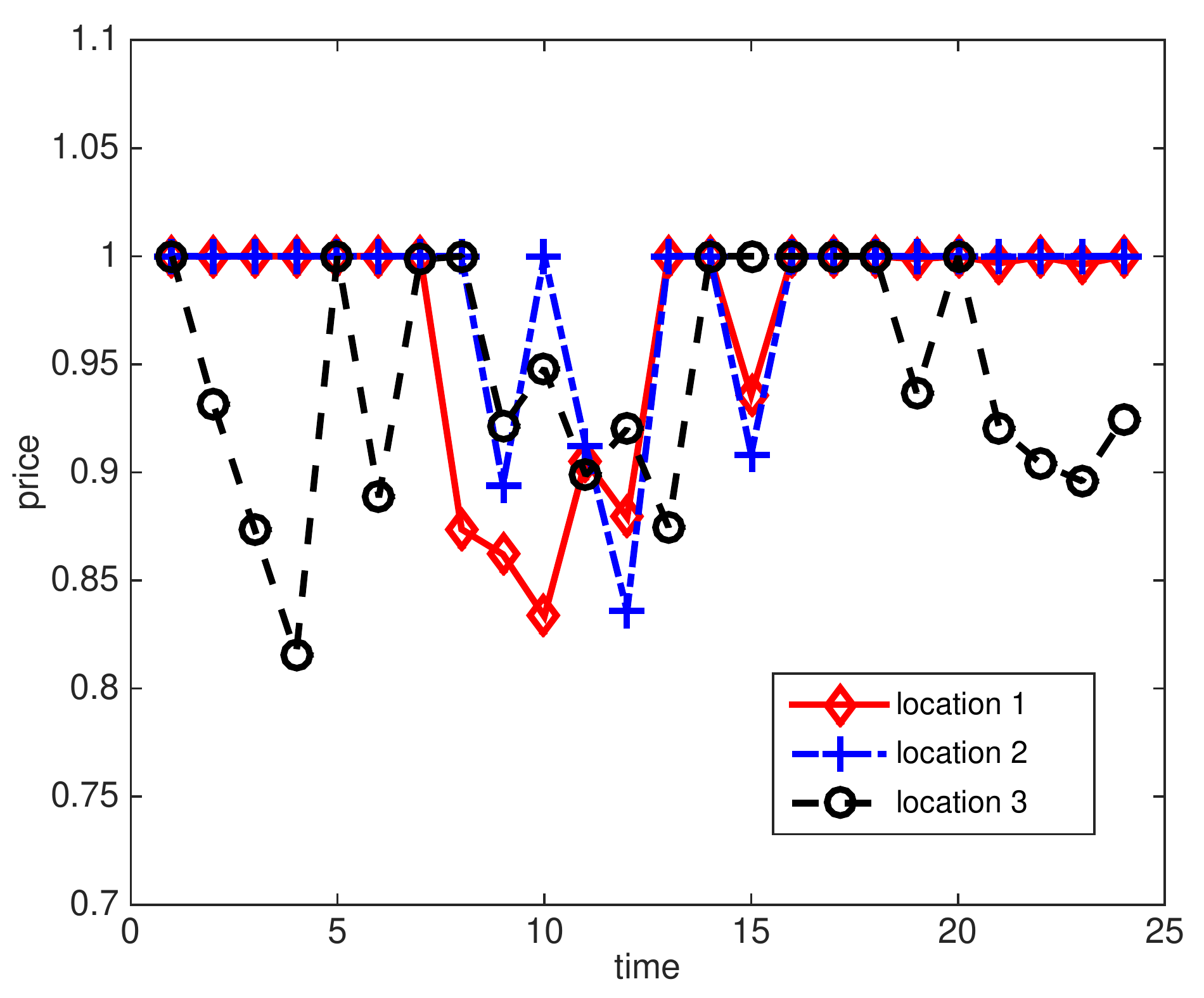}
  \caption{Optimized time and location dependent prices.}\label{fig:logprice}
  \end{minipage}
\vspace{-4mm}
\end{figure*}

\begin{figure*}
\vspace{-2mm}
\hspace{-5mm}
  \begin{minipage}[t]{0.33\linewidth}
  \centering
  \includegraphics[width=1\textwidth]{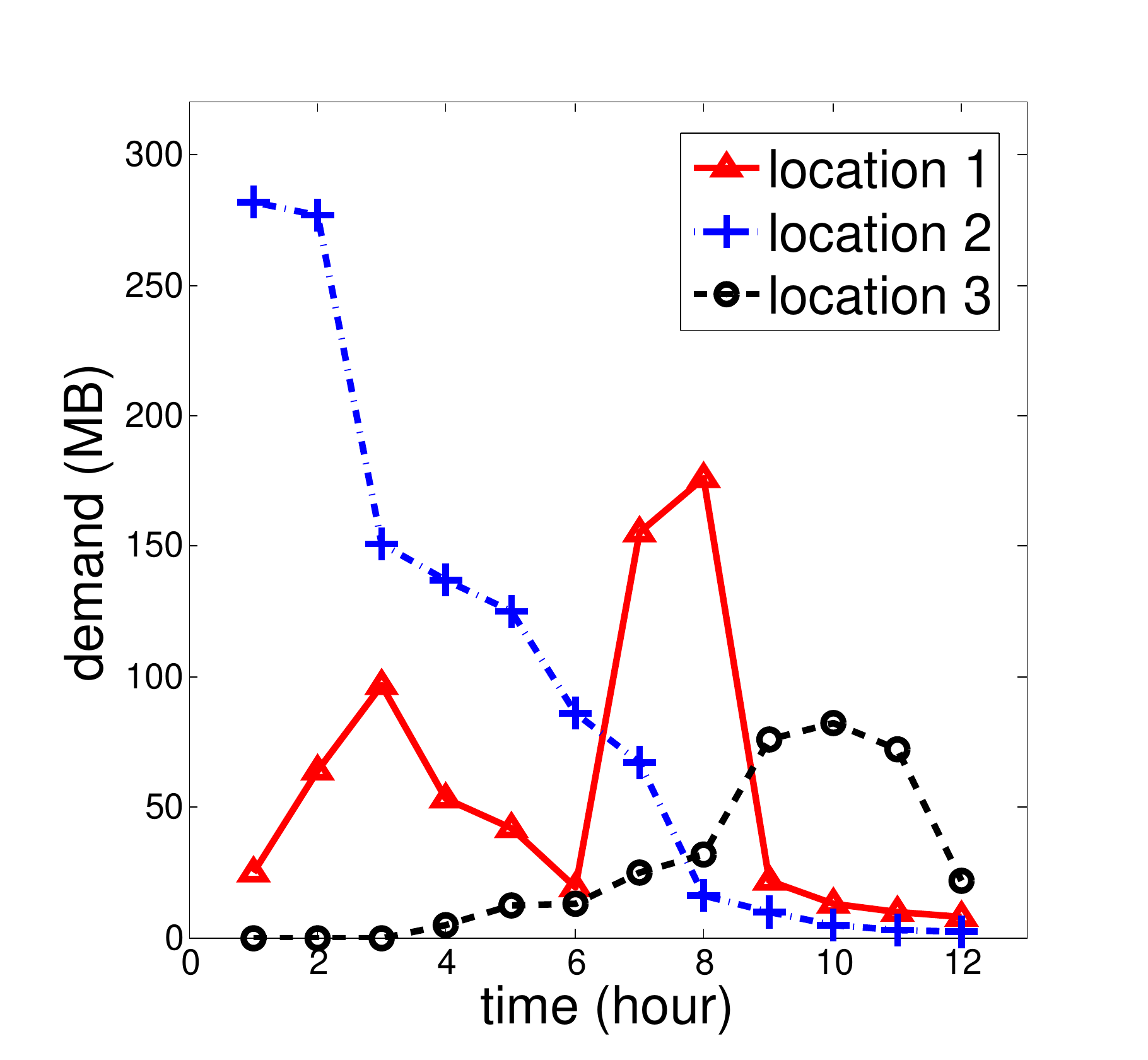}
  \caption{Initial traffic of 12 time slots.}\label{fig:ini}
  \end{minipage}
    \begin{minipage}[t]{0.32\linewidth}
      \centering
      \includegraphics[width=1\textwidth]{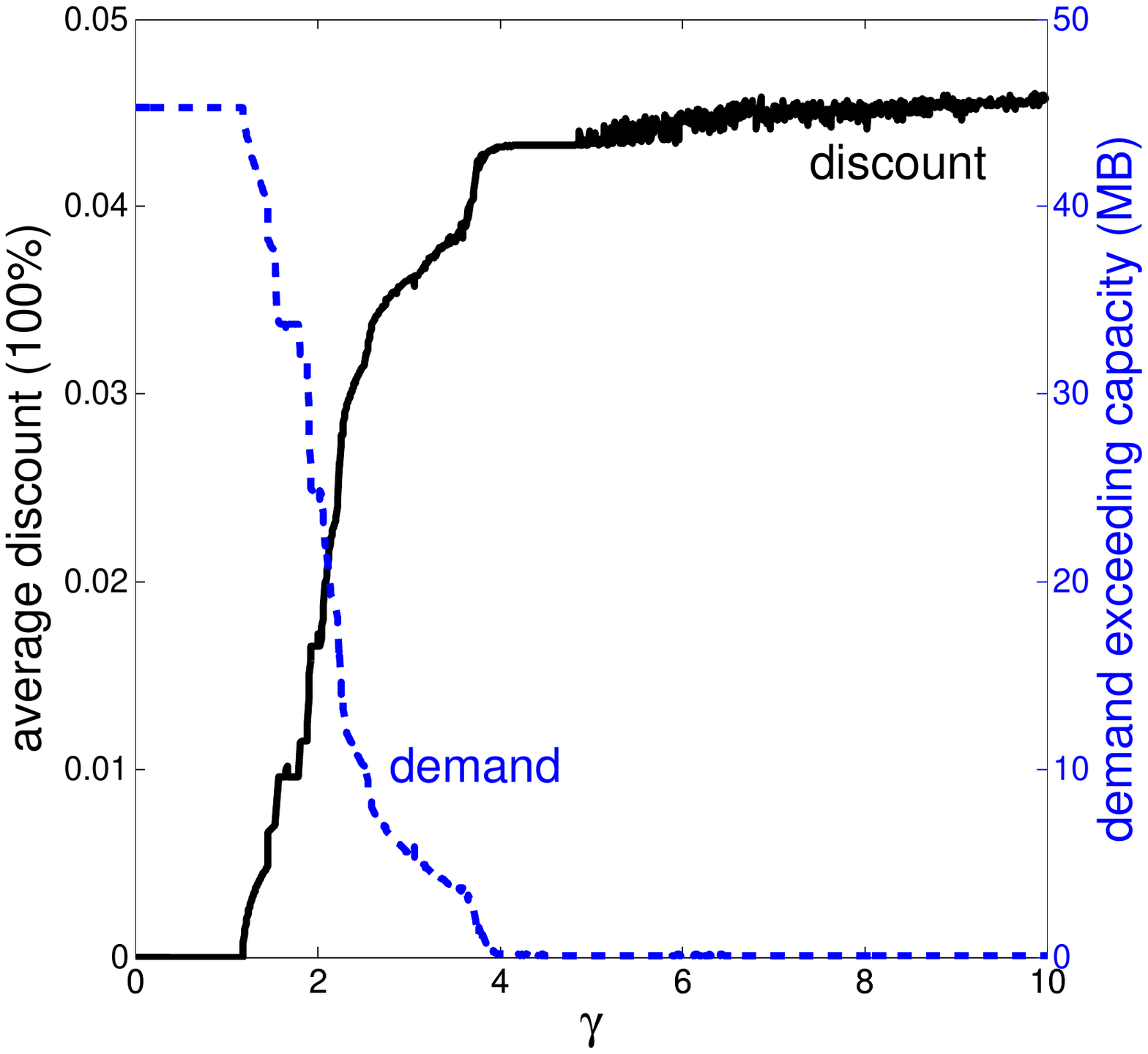}
      \caption{The impact of $\gamma$ with $\delta=0.8$.}\label{fig:gama2}
    \end{minipage}
  \begin{minipage}[t]{0.32\linewidth}
  \centering
  \includegraphics[width=1\textwidth]{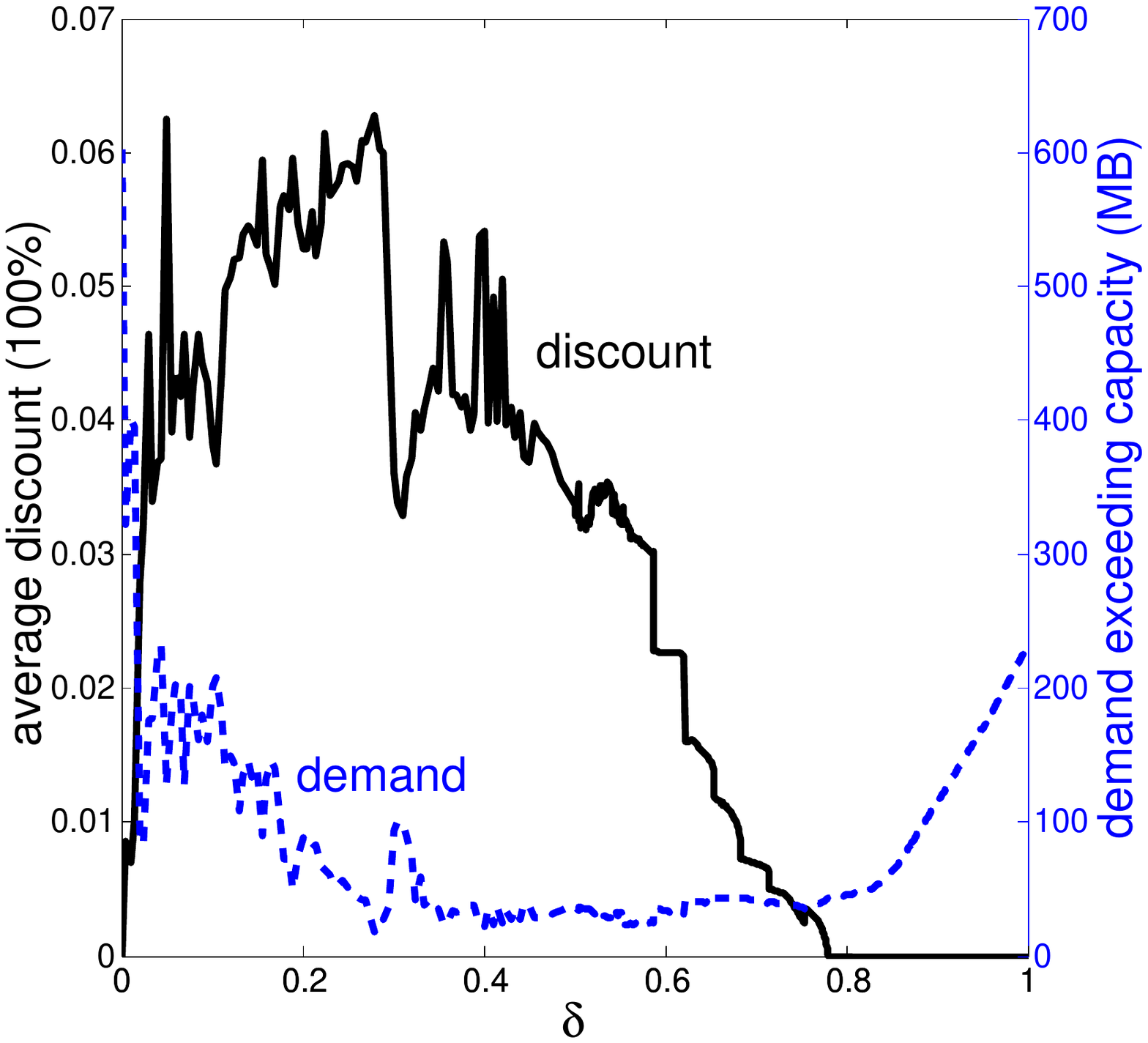}
  \caption{The impact of $\delta$ with $\gamma=1$.}\label{fig:delta2}
  \end{minipage}
\vspace{-2mm}
\end{figure*}

In this part, we first compare the performance of the SPG algorithm with that of the DYCORS algorithm. Then we use an example to illustrate the effectiveness of the proposed pricing scheme.

We compare the performance of our proposed SPG algorithm (Algorithm 2) with that of the DYCORS algorithm with $Nf_{\max}=2000$, by testing these two algorithms on some randomly generated instances. Table \ref{table-com} shows the detailed comparison. Simulation results show that the SPG algorithm can (almost) always find the global optimal solution\footnote{Notice that the SPG algorithm works slightly better than the DYCORS algorithm in terms of objective values. This is because the DYCORS algorithm does not use the gradient information and can only find an approximate solution in the neighborhood of the optimal solution with a given maximum number of function evaluations. The quality of the returned solution by the DYCORS algorithm depends on the maximum number of function evaluations. In contrast, the SPG algorithm uses the gradient information and can find a relatively better solution.}, with much less CPU time compared to the DYCORS algorithm.


To test the effectiveness our pricing scheme and gain more insights, we perform simulations based on real data from references \cite{Traffic} \cite{M2}. 
Figure \ref{fig:ini2} shows the initial traffic pattern under time and location independent pricing. The length of each time slot is $1$ hour.  
Each location corresponds to the coverage area of one base station. 
The data is measured in $10^8$ bytes. 
We assume that the network capacity $C=5$, the user's scheduling interval $T=12$, users' utility parameter $k_a=1$, and the time and location independent price benchmark $p_0=1$.

We set the operator's cost parameter $\gamma = 30$ and the users' delay parameter $\delta=0.6$. 
Figure \ref{fig:logusa} shows the aggregate shifted traffic under the optimized time and location dependent pricing. 
Figure \ref{fig:logprice} shows the corresponding optimized prices. 
The results in both figures are computed by the SPG algorithm. 



We can get some useful insights from the simulation results. First, the traffic can be smoothed by using our proposed time and location aware mobile data pricing scheme. In this example, the variance of traffic\footnote{The traffic variance is computed as $E\left[(x^{aft}(t,l)-E[x^{aft}(t,l)])^2\right]$.} is decreased by $62.65\%$. Second, our proposed pricing scheme leads to a win-win situation for both the operator and users. The cellular operator can decrease its total cost (consisting of the cost of demand exceeding capacity and the loss of revenue due to discounts) by $95.45\%$ (not directly shown in the figure). More specifically, the operator only uses $4.55\%$ of the initial network cost (under the time and location independent pricing) to provide price discounts to mobile users, and completely avoid the cost of demand exceeding capacity with the new optimized prices. Mobile users can increase their aggregate payoff (utility minus payment) by $55.37\%$ through proper traffic scheduling and taking advantage of the price discounts.

\subsubsection{The Impact of System Parameters}

We simulate and study how the average discount to the users and the total demand exceeding capacity change with the following two system parameters: the operator's cost parameter $\gamma$, and users' delay parameter $\delta$. The average discount to the users is defined as
$$\displaystyle \frac{\sum_{t=1}^{T_0} \sum_{l=1}^L \frac{p_0-p(t,l)}{p_0}}{T_0L},$$
and the total demand exceeding capacity is defined as
$$\sum_{t=1}^{T_0} \sum_{l=1}^L \max \{x(t,l)-C,0\}.$$
%

\begin{figure*}
\vspace{-5mm}
\hspace{-5mm}
    \begin{minipage}[t]{0.33\linewidth}
      \centering
      \includegraphics[width=1\textwidth]{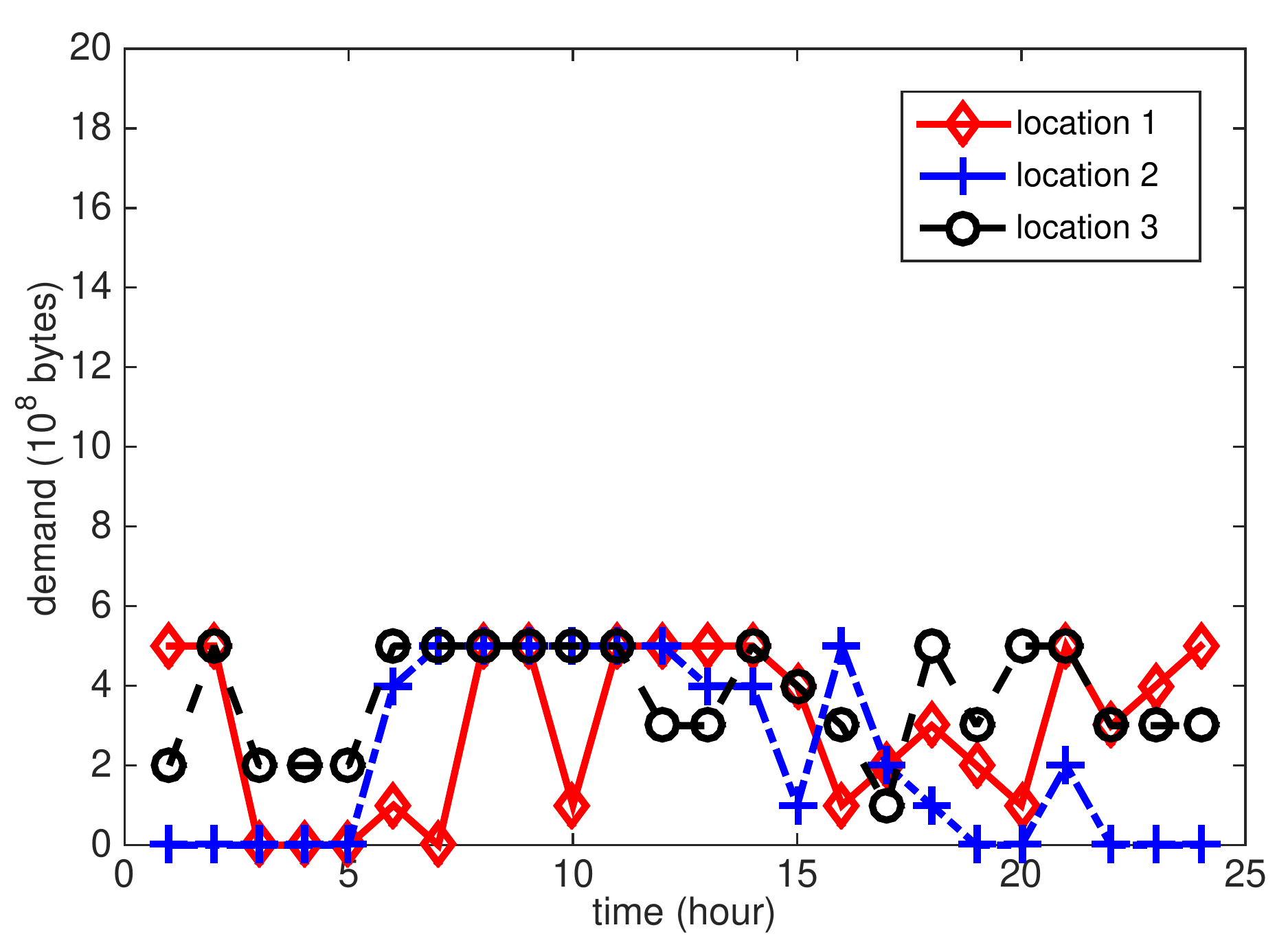}
      \caption{The usage under setting I ($\gamma=30$ and $\delta=0.8$)
.}\label{fig:usa1}
    \end{minipage}
  \begin{minipage}[t]{0.33\linewidth}
  \centering
  \includegraphics[width=1\textwidth]{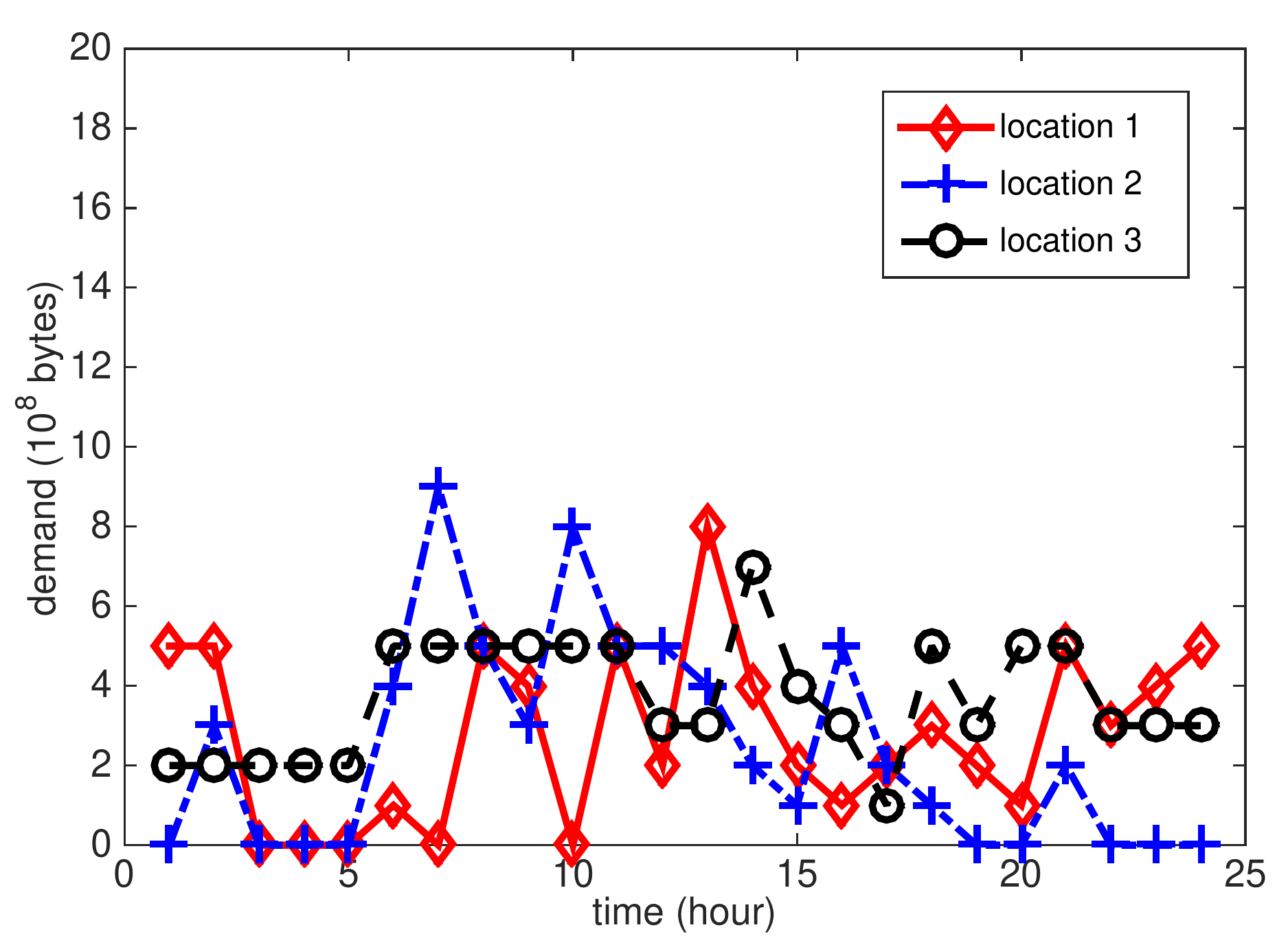}
  \caption{The usage under setting II ($\gamma=10$ and $\delta=0.6$).}\label{fig:usa2}
  \end{minipage}
  \begin{minipage}[t]{0.33\linewidth}
  \centering
  \includegraphics[width=1\textwidth]{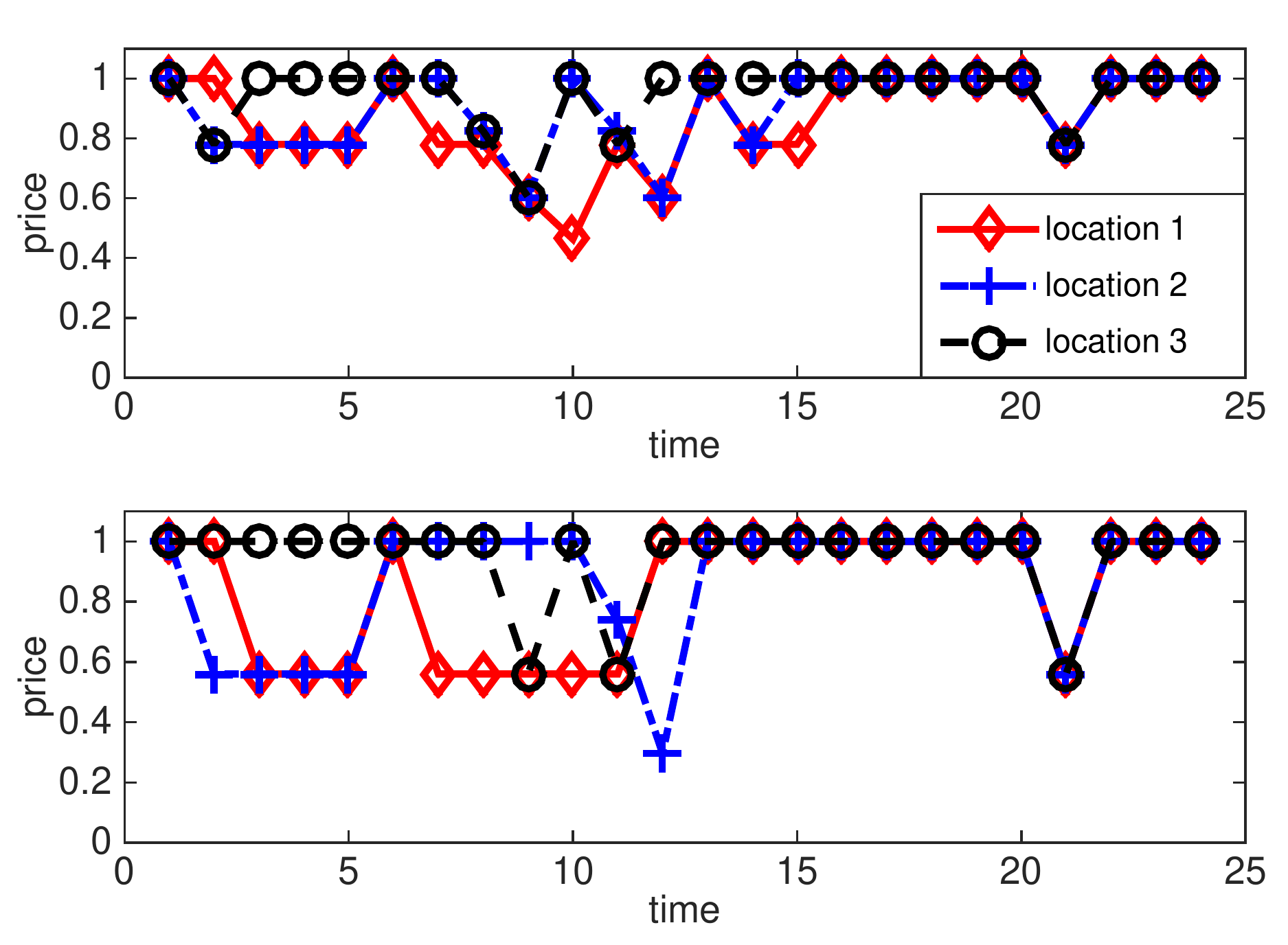}
  \caption{Prices under two settings (upper: setting I, lower: setting II).}\label{fig:price}
  \end{minipage}
\end{figure*}

Figure \ref{fig:ini} (whose dimension is smaller than that of Figure 3, due to computation complexity) shows the benchmark data traffic pattern under a time and location independent pricing scheme. We assume that the network capacity $C=100$ MB, users' scheduling interval $T=12$, the number of user type $A=1$, users' utility parameter $k_a=20$, and the benchmark price $p_0=1$.

We first analyze the impact of the cellular operator's cost parameter $\gamma$.

The cellular operator's goal is to minimize the total cost, which includes both the cost of demand exceeding capacity and loss of revenue due to discounts. When $\gamma$ increases, the operator has more incentives to offer deeper discounts to prevent demand exceeding capacity. Figure \ref{fig:gama2} illustrates how the average discount to the users (the black solid line) and the total demand exceeding capacity (the blue dashed line) change with $\gamma$, where users' delay parameter $\delta = 0.8$.

Figure \ref{fig:gama2} shows that the average discount offered by the operator is in general an increasing function of $\gamma$. As $\gamma$ becomes large, the operator has more incentive to offer a larger discount. When $\gamma$ is large enough, the operator has offered enough discount to the users to achieve the maximum smoothing effect, and any larger discount will not reduce the cost of demand exceeding capacity. Hence the discount eventually converges to a constant (around $0.046$ in Figure \ref{fig:gama2}).

Figure \ref{fig:gama2} also shows that the amount of demand exceeding capacity is a monotonically decreasing function of $\gamma$. As $\gamma$ becomes larger, a larger discount from the operator will reduce the amount of traffic exceeding the capacity. When $\gamma$ is large enough, the operator's average discount no longer changes, hence the traffic no longer changes and converges to a constant (in this case zero traffic exceeding capacity).

In summary, the operator tends to provide a large discount only when the demand exceeding capacity causes a large cost due to large value of $\gamma$.

Then we analyze the impact of users' delay parameter $\delta$.


A user's goal is to maximize his total payoff (utility minus payment). A larger $\delta$ means that the user is less sensitive to delay, and is more willing to delay his traffic to exploit the price discount. Figure \ref{fig:delta2} illustrates how the average discount (i.e., the black solid line) and the total demand exceeding capacity (i.e., the blue dashed line) change with $\delta$, where the operator's cost parameter $\gamma = 1$.

Figure \ref{fig:delta2} shows that the average discount first increases with $\delta$ (i.e., $\delta \in (0,0.3]$), then decreases (i.e., $\delta \in [0.3,0.8]$), and finally goes to zero (i.e., $\delta \in [0.8,1]$). When $\delta > 0$, it is possible to incentivize the user's behaviors by providing a discount. With $\delta$ increasing in the interval $(0,0.3]$, the operator will provide an increasingly larger discount to encourage users to shift the traffic, because a user is more willing to delay his traffic with a larger $\delta$ in this interval. With $\delta$ increasing in the interval $[0.3,0.8]$, a user becomes increasingly willing to shift his traffic even with a small discount, and hence the operator's optimal price discount actually decreases. When $\delta$ is large enough, because of the concavity of the logarithmic utility function, the user is willing to spread out his traffic in multiple time slots to maximize its utility even without a price incentive \cite{BehavioralEconomicsDUM}. Hence, when $\delta \in [0.8,1]$, there is no need for the operator to provide a discount (under the current cost parameter $\gamma = 1$).

Figure \ref{fig:delta2} also shows that the amount of demand exceeding capacity first decreases with $\delta$ (i.e., $\delta \in (0,0.3]$), then remains stable (i.e., $\delta \in [0.3,0.8]$), and finally increases (i.e., $\delta \in [0.8,1]$). With an increasing price discount for $\delta \in (0,0.3]$, the user is willing to delay more traffic which induces a smoother usage pattern. When $\delta \in [0.3,0.8]$, the traffic pattern remains almost the same as it has already been significantly flattened. When $\delta \in [0.8,1]$, because of the concavity of the logarithmic utility function, a user strongly prefers to delay his traffic to later time slots to maximize the utility, which may lead to newly created peak hours under the current parameter setting $\gamma=1$. Hence, with $\delta$ increasing in $[0.8,1]$, more demand is delayed to the later time slots which causes more demand exceeding capacity.

In summary, the operator only provides large price discounts when users are not willing to delay their traffic (i.e., the case where $\delta$ is small).

\subsection{Homogeneous Linear Utility Scenario}

\subsubsection{The Effectiveness of Our Pricing Scheme}

In this part, we verify the effectiveness of the proposed pricing scheme.\footnote{Numerical results show that the BCD algorithm can achieve global optimality for small-scale problems, comparing with a benchmark branch and bound algorithm. Such a comparison becomes infeasible when the problem size becomes large, due to the high complexity of the branch and bound algorithm.}

We use the same network example shown in Figure \ref{fig:ini2} to illustrate the effectiveness of the BCD algorithm.

We simulate two different settings of parameters. Under setting I, we set $\gamma = 30$ and $\delta=0.95$, and our proposed time and location aware pricing (computed by the BCD algorithm in Algorithm 3) leads to a shifted traffic pattern as shown in Figure \ref{fig:usa1}. Under setting II, we set $\gamma=10$ and $\delta=0.7$, and the corresponding shifted traffic pattern is shown in Figure \ref{fig:usa2}. Figure \ref{fig:price} shows the operator's optimal time and location aware prices for both settings, in which the upper one corresponds to Figure \ref{fig:usa1}, and the lower one corresponds to Figure \ref{fig:usa2}.

We can get some useful observations from the simulation results. First, how much traffic can be smoothed heavily depends on the system parameters. Under Setting I where $\gamma=30$ and $\delta=0.8$, the variance for data usage $\left\{x(t,l)\right\}$ is decreased by $56.39\%$. Under Setting II where $\gamma=10$ and $\delta=0.6$, the variance for data usage is decreased by $46.11\%$. 
This is because a larger cost parameter $\gamma$ makes the operator more willing to provide price incentives, and a larger delay tolerance parameter $\delta$ makes the users more willing to delay the traffic. 
Second, our proposed pricing scheme leads to a win-win situation for both the operator and users. The cellular operator can decrease its total cost by $98.01\%$ under setting I and $70.20\%$ under setting II. Mobile users can increase their payoff by $107.37\%$ under setting I and $102.80\%$ under setting II.

\subsubsection{The Impact of System Parameters}

We use the same network example showed in Figure \ref{fig:ini} to analyze the impact of the operator's cost parameter $\gamma$ and users' delay parameter $\delta$. 

We first analyze the impact of the cellular operator's cost parameter $\gamma$.

\begin{figure}
  \centering
  \includegraphics[width=0.35\textwidth]{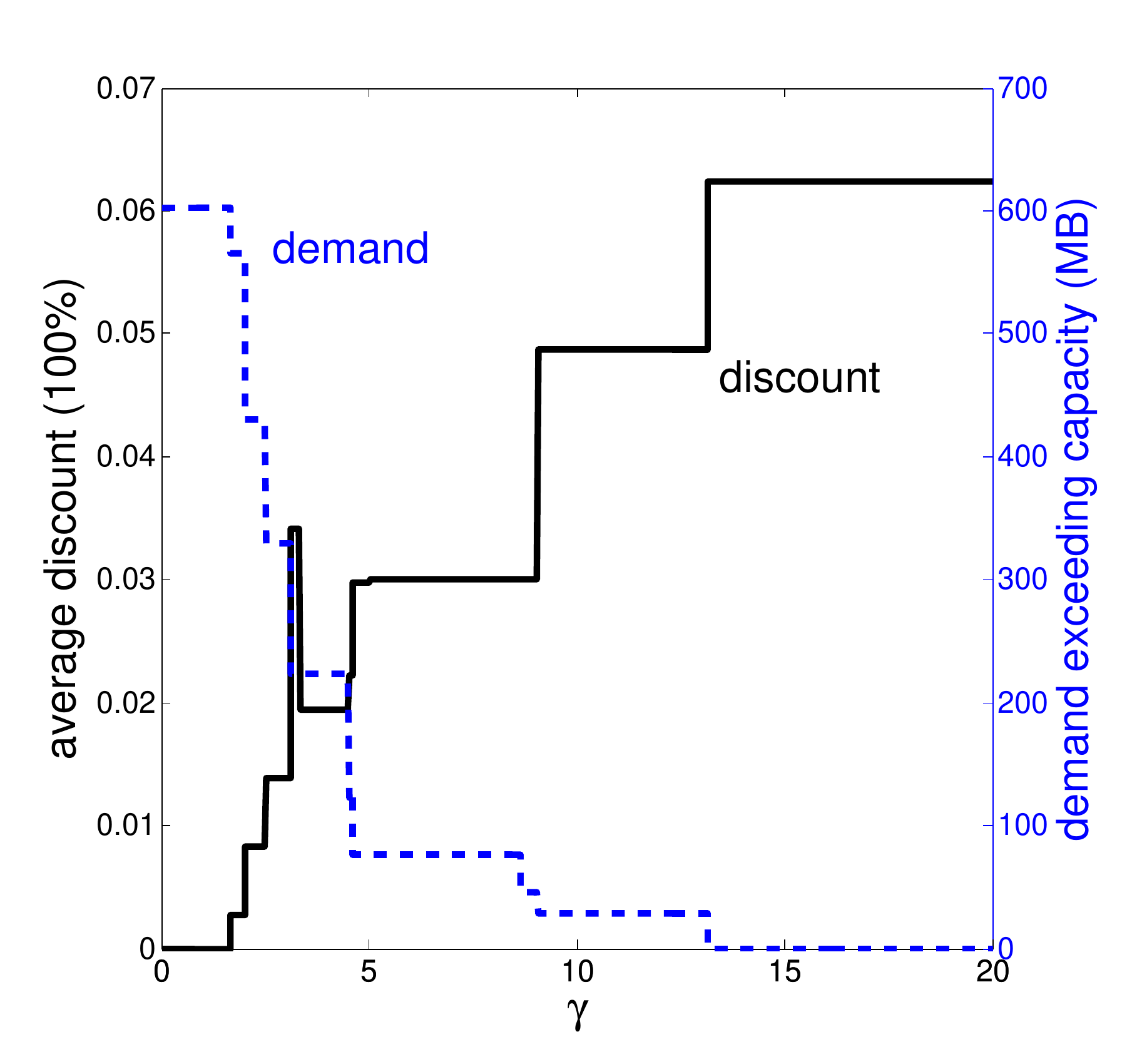}
  \vspace{-3mm}
  \caption{The impact of $\gamma$ with $\delta=0.9$.}\label{fig:gama}
\end{figure}

Figure \ref{fig:gama} illustrates how the average discount to the users (i.e., the black solid line) and the total demand exceeding capacity (i.e., the blue dashed line) change with $\gamma$, where users' delay parameter $\delta = 0.9$.

Figure \ref{fig:gama} shows that the average discount offered by the operator is in general an increasing step function of $\gamma$. It is increasing, as a larger cost $\gamma$ encourages the operator to give a larger discount. It is a step function, because a user's scheduling decision is not continuous with respect to the price.
There is an exception when $\gamma = 3.4$, where the average discount decreases instead of increases with $\gamma$. The reason is that under this particular parameter setting, users choose to shift the traffic to immediate
adjacent time slots rather than slots at even later times, which leads to a smaller discount.

Figure \ref{fig:gama} also shows that the amount of demand exceeding capacity is a monotonically decreasing function of $\gamma$. As $\gamma$ becomes larger, a larger discount from the operator will reduce the amount of traffic exceeding the capacity.

\begin{figure}
  \centering
  \includegraphics[width=0.35\textwidth]{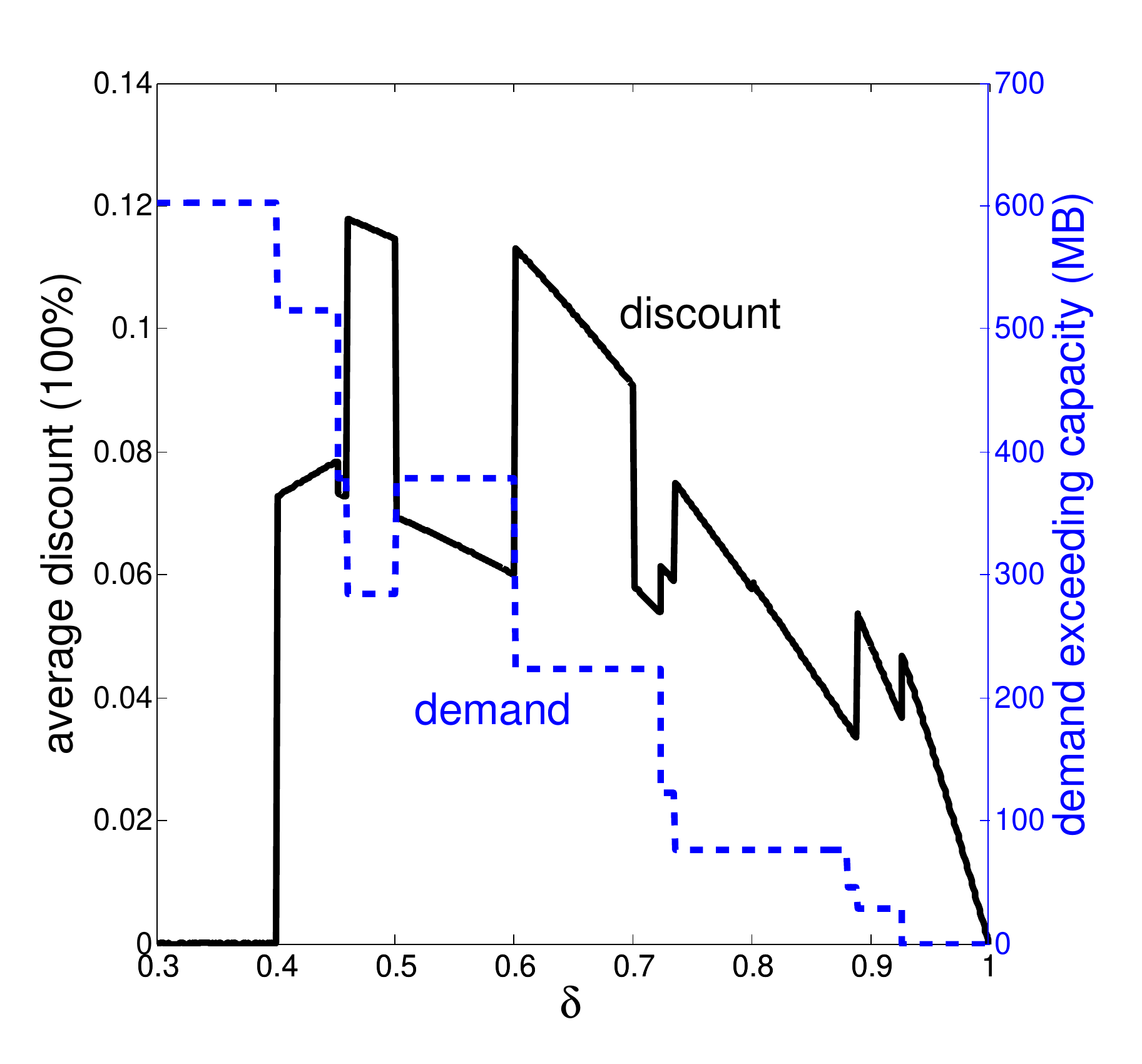}
  \vspace{-3mm}
  \caption{The impact of $\delta$ with $\gamma=10$.}\label{fig:delta}
\end{figure}

Then we analyze the impact of users' delay parameter $\delta$.

Figure \ref{fig:delta} illustrates how the average discount (i.e., the black solid line) and the total demand exceeding capacity (i.e., the blue dashed line) change with $\delta$, where the operator's cost parameter $\gamma = 10$.

Figure \ref{fig:delta} shows that the average discount is a piecewise function of $\delta$, where most of the segments are decreasing in $\gamma$ (e.g., the segment corresponding to [0.6,0.7]). The reason is that a segment corresponds to the same shifted usage pattern, and a larger $\delta$ reduces the operator's needs of providing price discounts. However, for the general trend of the average discount, it first increases with $\delta$ (i.e., $\delta \in [0.3,0.45]$)
and then decreases with $\gamma$ (i.e., $\delta \in [0.45,1]$). When $\delta$ is very small, there is no point of providing any discount to users, as users are very unlikely to delay traffic. When $\delta$ increases to the point that it is possible to incentive users' behaviors, there is a sharp increase of the discount. When $\delta$ is large enough, the operator only needs to provide a small discount to incentivize the users to shift to the desirable time slot.

Figure \ref{fig:delta} also shows that the amount of demand exceeding capacity is a decreasing function of $\delta$. There is a special case when $\delta = 0.5$. The main reason is that the integrated problem not only considers users' scheduling problem, but also considers the operator's pricing problem. When $\delta = 0.5$, we can see from the discount curve in Figure \ref{fig:delta} that there is a sharp decrease of the offered discount, which leads to a relatively small increase in the demand exceeding capacity.

In summary, the logarithmic case and the linear case are different. In the linear case, the objective function of the operator is discontinuous with respect to the prices; while it is continuous in the logarithmic utility case. As a result, the operator's optimal pricing and users' scheduling traffic are not smooth under the linear utility function, while they are continuous under the logarithmic utility function. These are consistent with our analysis in Sections \ref{sec:homo1} and \ref{sec:homo2}.

\section{Conclusion}\label{sec:conc}

In this paper, we study the time and location aware pricing scheme for wireless mobile data networks. We use a two-stage decision process to model the interaction between the cellular operator and users. Simulation results show that our time and location aware pricing scheme not only reduces the operator's cost but also increases users' payoff. We further derive some insights for industry practice. The operator should provide price discounts at less crowded time slots and locations to incentivize users to smooth their traffic. The operator will only provide deep discount when the cost of serving demand exceeding capacity is high, or when the users' willingness to delay traffic is low.

Our next step plan is to conduct large scale comprehensive simulation studies of the algorithm performance based on realistic mobile data usage traces, and create mobile apps to further help users make automated traffic scheduling decisions.
The integration of WiFi and cellular (e.g., 4G) networks is also a promising topic for future exploration.

\appendices

\section{Robust Optimization Considering Errors in the Mobility Prifiles}

We discuss the accuracy of the mobility profiles and the impact of errors to the optimization. 

First, Ghosh \emph{et al.} claimed that the profile-based location predictions are more accurate than a common statistical approach. 
Essentially, the mobility profiles take into account the daily routes of mobile users. 
Furthermore, the methods proposed in \cite{M2} can filter out noises (i.e., very brief location stay durations) to achieve more accurate characterization of mobility profiles.

We then discuss the impact of the errors to the optimization.
If the mobility profiles have errors, then these errors will affect both the objective function and constraint of Problem 1. 
To highlight the dependence of the objective function of Problem 1 on the mobility profiles, we write the objective function of Problem 1 as $U(\bx_a(t,l),\boldsymbol{\beta}^a)- P(\bx_a(t,l),\boldsymbol{\beta}^a).$

To deal with the errors in the objective, we can use the idea of robust optimization. 
Let $\Phi^a$ be the uncertainty set of $\boldsymbol{\beta}^a.$ 
Without loss of generality, we can model it as an ellipsoid intersecting with a probability simplex. 
Instead of maximizing the objective of Problem 1, we can maximize the optimal value of problem
\begin{equation}\label{p1}
\min_{\boldsymbol{\beta^a}\in\Phi^a} U(\bx_a(t,l),\boldsymbol{\beta}^a)- P(\bx_a(t,l),\boldsymbol{\beta}^a).
\end{equation} 

To deal with the errors in the constraint, we introduce a tolerance parameter $\epsilon>0$ and enforce
\begin{align}\label{p2}
& \left|x_a(t,l|t,l)+\sum_{t'=t+1}^{T_t}\sum_{l'=1}^L\beta_a(t',l'|t,l)x_a(t',l'|t,l)-x_a^{ini}(t,l)\right| \notag \\ 
&\leq \epsilon,~\forall~\boldsymbol{\beta}^a\in \Phi^a.
\end{align}
Note that the introduction of $\epsilon>0$ in the above is necessary, since it is generally not possible for some $\bx_a(t,l)$ to satisfy the linear constraint $$x_a^{ini}(t,l)=x_a(t,l|t,l)+\sum_{t'=t+1}^{T_t}\sum_{l'=1}^L\beta_a(t',l'|t,l)x_a(t',l'|t,l)$$ for all $\boldsymbol{\beta}^a\in \Phi^a$ (unless $\Phi^a$ contains only a single point).

Therefore, when there are errors in the mobility profiles, we can solve the following robust optimization problem (instead of Problem 1)
\begin{equation}\label{problemrobust}
\begin{array}{cl}
\displaystyle \max_{\bx^a(t,l) \geq \boldsymbol{0}} & \displaystyle \min_{\boldsymbol{\beta^a}\in\Phi^a} U(\bx_a(t,l),\boldsymbol{\beta}^a)- P(\bx_a(t,l),\boldsymbol{\beta}^a)\\[5pt]
\mbox{s.t.} & \Big|x_a(t,l|t,l)+\displaystyle\sum_{t'=t+1}^{T_t}\sum_{l'=1}^L\beta_a(t',l'|t,l)x_a(t',l'|t,l) \\
& -x_a^{ini}(t,l)\Big|\leq \epsilon,~\forall~\beta^a\in \Phi^a.
\end{array}
\end{equation}
When there are no errors in the mobility profiles, we can set $\epsilon=0$ in Problem \eqref{problemrobust} and it reduces to Problem 1.

Problem \eqref{problemrobust} is a semi-infinite programming problem \cite{Semiinfinite}, since it involves a finite number of variables and an infinite number of constraints. 
Although it is hard to solve Problem \eqref{problemrobust} analytically, we can numerically solve it, for instance, by the DYCORS algorithm. 
This is because given any $\bx_a(t,l),$ we can compute its objective value and check its feasibility efficiently. 
Specifically, computing the objective value of Problem \eqref{problemrobust} is equivalent to solving convex problem \eqref{p1}. 
Checking whether $\bx_a(t,l)$ is feasible to Problem \eqref{problemrobust} requires solving the following problem:  
\begin{equation}
\begin{array}{cl}
\displaystyle \max_{\boldsymbol{\beta^a}\in\Phi^a} & |x_a(t,l|t,l)+ \displaystyle  \sum_{t'=t+1}^{T_t}\sum_{l'=1}^L\beta_a(t',l'|t,l)x_a(t',l'|t,l)\\[5pt]
& -x_a^{ini}(t,l)|
\end{array}
\end{equation}
The above problem can be solved by solving two convex problems
$$\min_{\boldsymbol{\beta^a}\in\Phi^a}-\sum_{t'=t+1}^{T_t}\sum_{l'=1}^L\beta_a(t',l'|t,l)x_a(t',l'|t,l)$$
and 
$$\min_{\boldsymbol{\beta^a}\in\Phi^a}\sum_{t'=t+1}^{T_t}\sum_{l'=1}^L\beta_a(t',l'|t,l)x_a(t',l'|t,l).$$

The key conclusion in our paper will not change if the errors are relatively small. 
Specifically, the operator will set higher prices at peak hours and crowded locations. 
However, the mobility patterns may affect the operator's decision of when and where to provide the lower prices, and how much to reduce the prices.

\section{An Example of Discontinuous $H(\boldsymbol{p})$}\label{app1}

\begin{table}[t]
\caption{\textsc{Setup of the Example}} \label{table-example}
\centering
\begin{tabular}[h]{|c|c|}
  \hline\hline
{Parameters} &  {Values}
\\\hline
        {Total time slots}  &   {$T_0=2$}
\\\hline
        {Total locations} &  {$L=1$}\\
\hline {Number of user types} & $A=1$\\
\hline Scheduling time interval & $T=2$\\
\hline Global mobility profile & $\boldsymbol{\alpha}=[1 ~1]^T$\\
\hline Local mobility profile & $\boldsymbol{\beta}=[1 ~1]^T$\\
\hline Unit cost for demand exceeding capacity & $\gamma=1$\\
\hline Initial data traffic pattern &$\boldsymbol{x}^{ini}=[1 ~1]^T$\\
\hline Network capacity & $C=1$\\
\hline Delay parameter & $\delta=1$\\
\hline\hline
\end{tabular}
\end{table}

We consider a linear utility function $u_a(x)=x$. In the setup shown in Table \ref{table-example}, we evaluate $H(\boldsymbol{p})$ at two different prices $\boldsymbol{p}_1=[1 ~1-\varepsilon]^T$ and $\boldsymbol{p}_2=[1-\varepsilon ~1]^T,$ where $\varepsilon$ is a sufficient small positive number. When $\boldsymbol{p}=\boldsymbol{p}_1,$ a user will schedule his traffic as $\boldsymbol{x}^{aft}=[0 ~2]^T$ to maximize his payoff, and $H(\boldsymbol{p}_1)=2 \varepsilon -1$; while when $\boldsymbol{p}=\boldsymbol{p}_2$, a user's optimal traffic (to maximize his payoff) is $\boldsymbol{x}^{aft}=[1 ~1]$, and $H(\boldsymbol{p}_2)=\varepsilon-2$. Hence, a small change of $\boldsymbol{p}$ ($\triangle \boldsymbol{p}=\boldsymbol{p}_1-\boldsymbol{p}_2=[\varepsilon ~-\varepsilon]^T$) leads to a large change of $H(\boldsymbol{p})$ ($\triangle H = H(\boldsymbol{p}_1) - H(\boldsymbol{p}_2)=1 + \varepsilon$ ). This implies that the objective function $H(\boldsymbol{p})$ of Problem 4 can be discontinuous (and hence nondifferentiable) with respect to $\boldsymbol{p}$.

\section{The Impact of $\epsilon$ on the Algorithm Performance}

\begin{figure*}
  \begin{minipage}[t]{0.33\linewidth}
  \centering
  \includegraphics[width=1\textwidth]{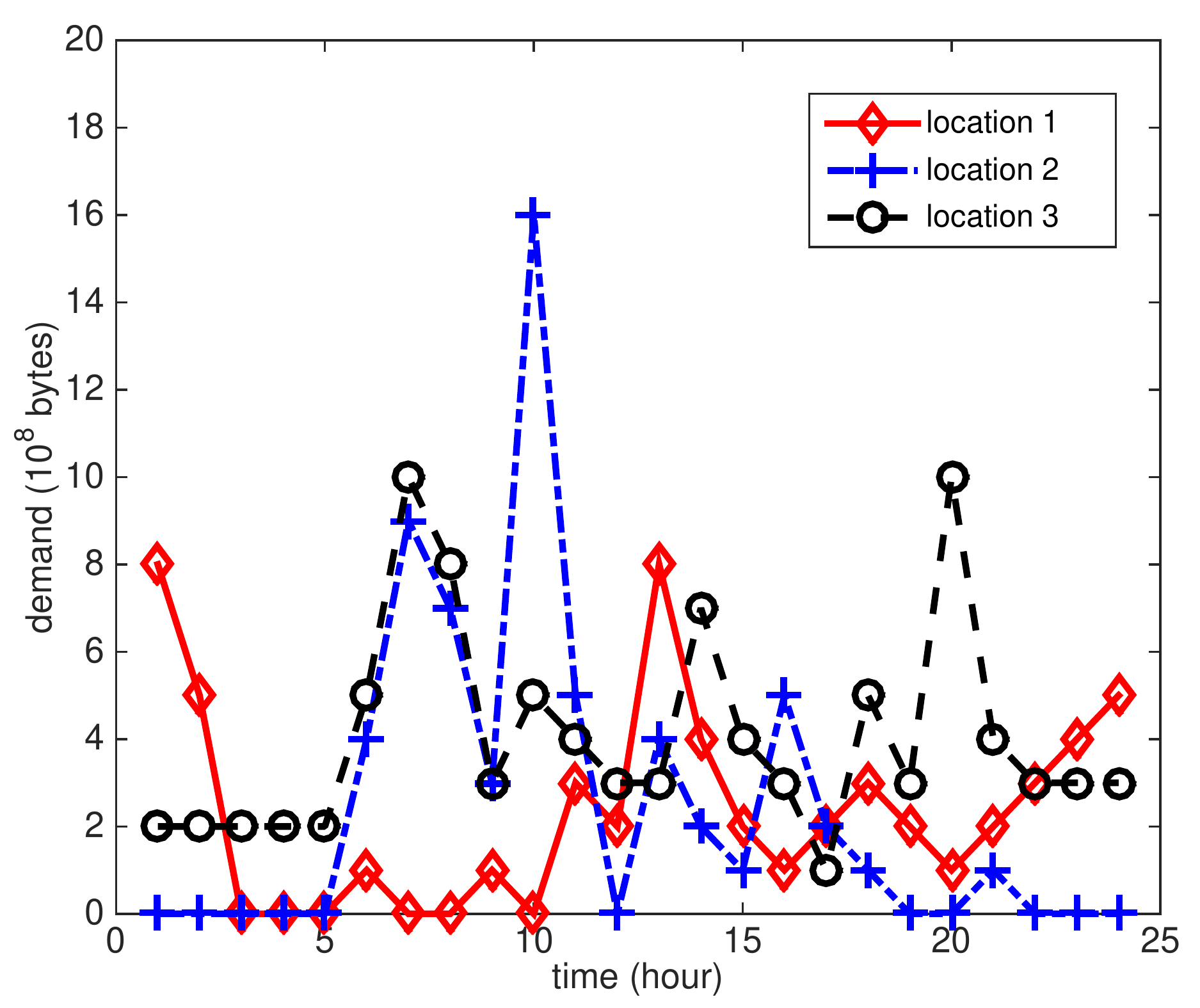}
  \caption{Initial traffic of 24 time slots}\label{fig:ini2}
  \end{minipage}
    \begin{minipage}[t]{0.33\linewidth}
      \centering
      \includegraphics[width=0.98\textwidth]{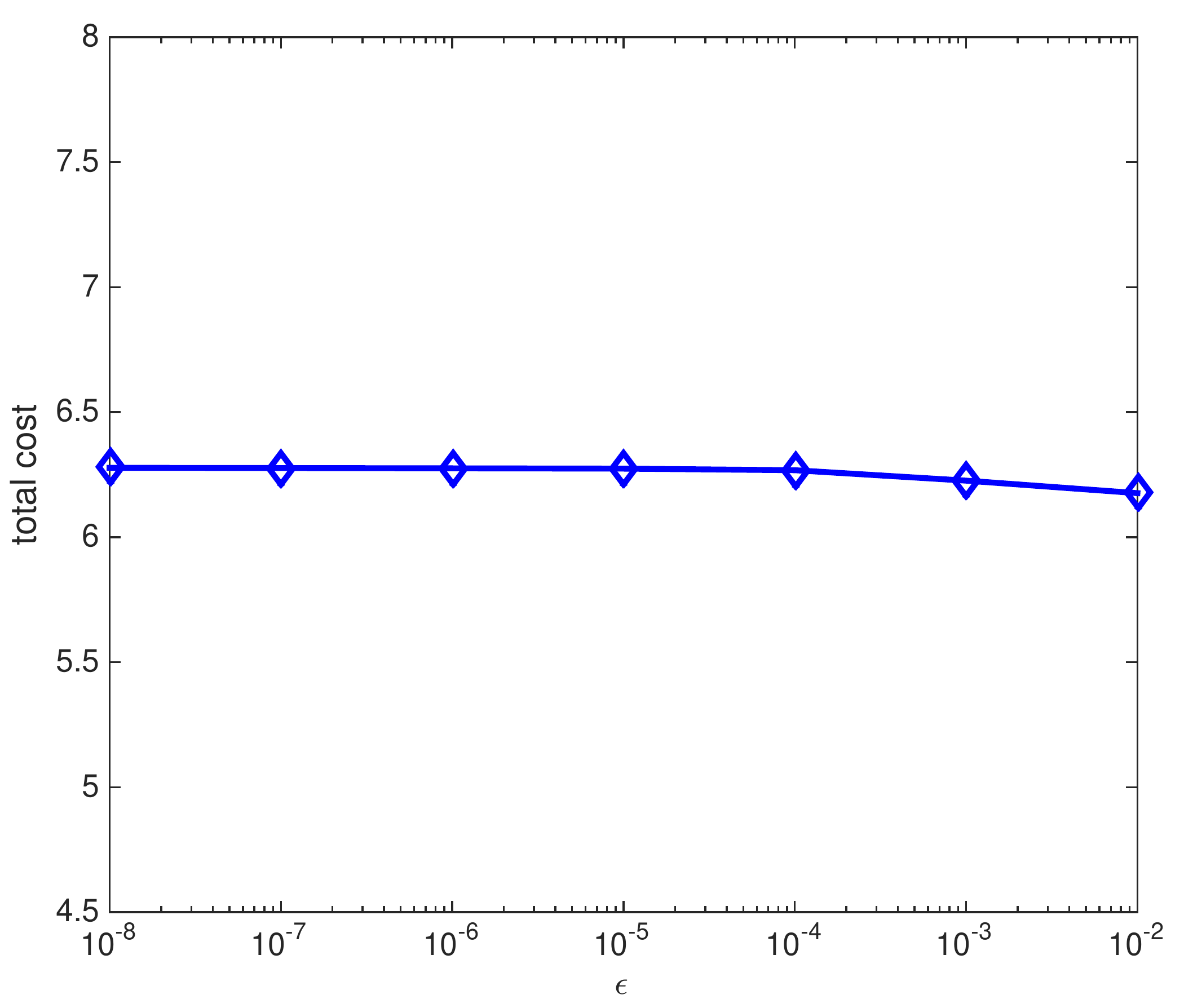}
      \caption{Total cost under different $\epsilon$}\label{fig:Impact}
    \end{minipage}
  \begin{minipage}[t]{0.33\linewidth}
  \centering
  \includegraphics[width=1.01\textwidth]{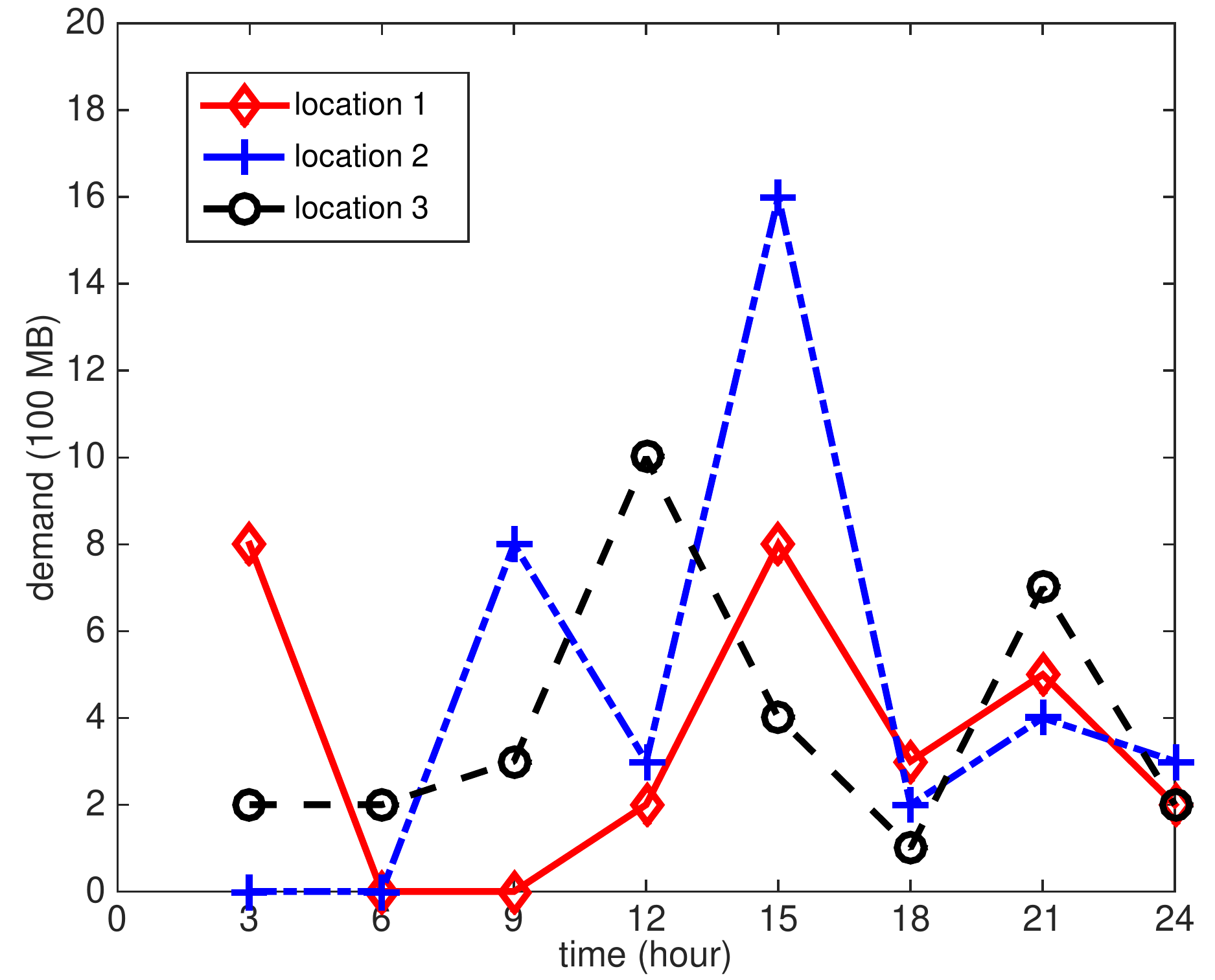}
  \caption{Initial data usage under flat rate pricing}\label{fig:InitialDataTDP}
  \end{minipage}
\end{figure*}

We first introduce the general idea of binary search. 
We take the binary search for $\lambda_a^\ast(t,l)$ as an example, with the initial search interval $[\lambda_a^L, \lambda_a^U]$, where $\lambda_a^L=\underline \lambda_a(t,l)$ and $\lambda_a^U= \bar \lambda_a(t,l)$. 
The general idea of the binary search is to evaluate the value of function $g(\cdot)$ at the middle point of the search interval, i.e., $\lambda_a^M=\frac{\lambda_a^L+\lambda_a^U}{2}$. 
Here function $g(\cdot)$ is:
\begin{align}
g(\lambda_a (t,l))&=x_a(t,l|t,l)+ \sum_{t'=t+1}^{T_t} \sum_{l'=1}^L \beta_a(t',l'|t,l) x_a(t',l'|t,l) \notag \\
&-x_a^{ini}(t,l), \label{eq:g}
\end{align}
where
\begin{align}
x_a(t,l|t,l)&=\max \left\{\frac{k_a}{p(t,l)+\lambda_a (t,l)}-1,0\right\}, \label{eq:Z1}\\
x_a(t',l'|t,l)&=\max \left\{\frac{k_a \delta_a^{t'-t}}{p(t',l')+\lambda_a (t,l)}-1,0\right\}. \label{eq:Z2}
\end{align}
If the function value at the middle point is equal to the target value, i.e., $g(\lambda_a^M)=0$, then the iteration ends and the optimal solution is the middle point. 
Otherwise, we replace the upper bound or lower bound of the search interval with the middle point, hence shorten the length of the search interval by half. 
The iteration process continues until the optimal solution is found or the search interval is small enough. 
After $n$ iterations, the length of the search interval is $\frac{\overline{\lambda}_a(t,l)-\underline{\lambda}_a(t,l)}{2^n}$.
Mathematically, the process ends when the length of the search interval is small enough, i.e., 
\[ \frac{\overline{\lambda}_a(t,l)-\underline{\lambda}_a(t,l)}{2^n} \leq \epsilon ,\]
where $\epsilon$ is a small enough positive number. 
When the iteration process ends, we choose the middle point as the solution, i.e., 
\[\lambda_a(t,l)=\frac{\lambda_a^L+\lambda_a^U}{2}.\] 
To find an $\epsilon$-optimal solution, the binary search needs $\lceil \log_2(\frac{\overline{\lambda}_a(t,l)-\underline{\lambda}_a(t,l)}{\epsilon}) \rceil$ iterations. 
Hence, the complexity of binary search is $\mathcal{O} (\log \frac{1}{\epsilon})$, which is polynomial in terms of $\frac{1}{\epsilon}$.

Now we discuss the error of the $\epsilon$-optimal $\lambda_a(t,l)$ generated by the binary search. 
If the target value is equal to the function value at the middle point, then the middle point is the optimal solution, and the error is zero. 
Otherwise, the error of the $\epsilon$-optimal $\lambda_a(t,l)$ is \cite{BinarySearch, Bisection}:
\[\Delta := |\lambda_a(t,l)-\lambda_a^*(t,l)|\leq \frac{\overline{\lambda}_a(t,l)-\underline{\lambda}_a(t,l)}{2^{n+1}}\leq \frac{1}{2}\epsilon.\]
In our simulations, we set $\epsilon = 10^{-6}$.


To check the impact of $\epsilon$ on the performance of the SPG algorithm, we perform simulations on the example in Section 4.1.1, and Figure \ref{fig:ini2} shows the initial traffic pattern under time and location independent pricing of the example. 
The length of each time slot is $1$ hour.  
Each location corresponds to the coverage area of one base station. 
The data is measured in $10^8$ bytes. 
We assume that the network capacity $C=5$, the user's scheduling interval $T=12$, users' utility parameter $k_a=1$, and the time and location independent price benchmark $p_0=1$.
We set the operator's cost parameter $\gamma = 30$ and the users' delay parameter $\delta=0.6$.

We calculate the operator's total cost using the SPG algorithm under different values of $\epsilon$, i.e., $\epsilon = 10^{-2}$, $\epsilon = 10^{-3}$, $\epsilon = 10^{-4}$, $\epsilon = 10^{-5}$, $\epsilon = 10^{-6}$, $\epsilon = 10^{-7}$, and $\epsilon = 10^{-8}$. 
Figure \ref{fig:Impact} shows the optimal total cost obtained by the SPG algorithm under different $\epsilon$. 
Simulation results show that as long as $\epsilon$ is not too large (i.e., $\epsilon \leq 10^{-4}$), different choices of $\epsilon$ will not affect the final results significantly.


\section{Derivations of Upper and Lower Bounds of $\lambda_a^{\ast} (t,l)$}\label{app2}

\begin{figure*}
\hspace{-5mm}
    \begin{minipage}[t]{0.25\linewidth}
      \centering
      \includegraphics[width=0.96\textwidth]{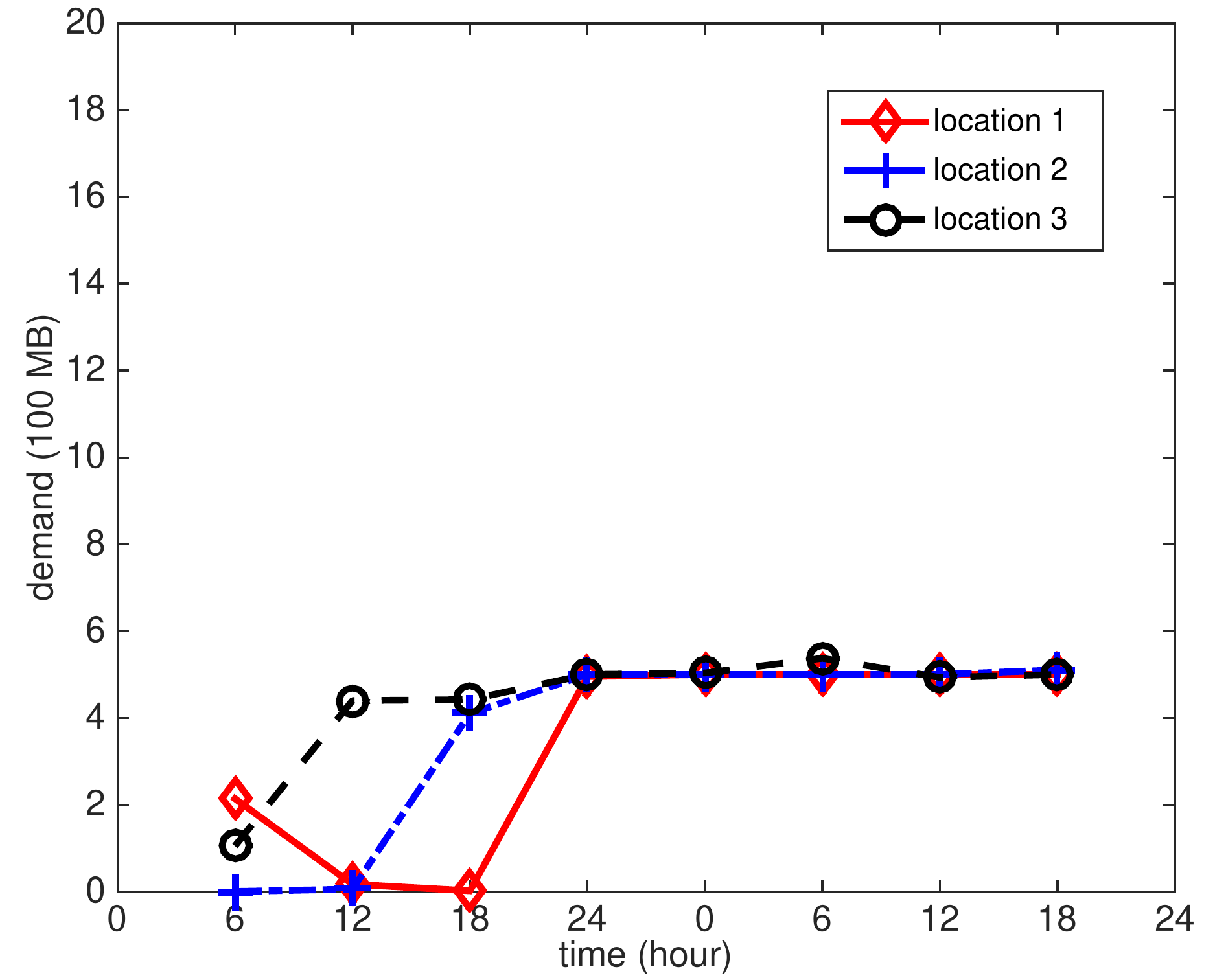}
      \caption{Usage under time and location aware pricing (the logarithmic utility case)}\label{fig:NewDataTLPSPG}
    \end{minipage}
  \begin{minipage}[t]{0.25\linewidth}
  \centering
  \includegraphics[width=1\textwidth]{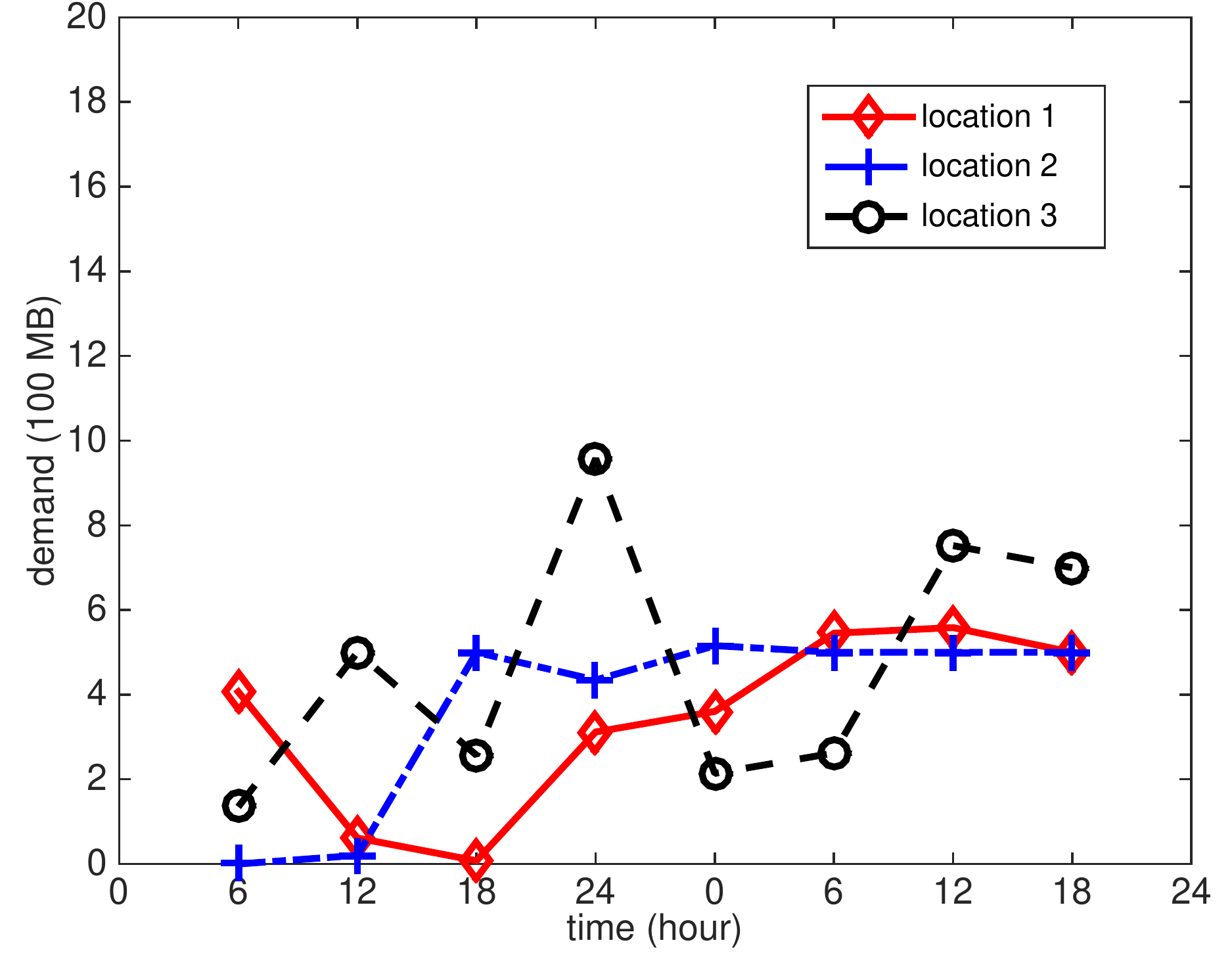}
  \caption{Usage under time-dependent pricing (the logarithmic utility case)}\label{fig:NewDataTDPSPG}
  \end{minipage}
    \begin{minipage}[t]{0.25\linewidth}
      \centering
      \includegraphics[width=1\textwidth]{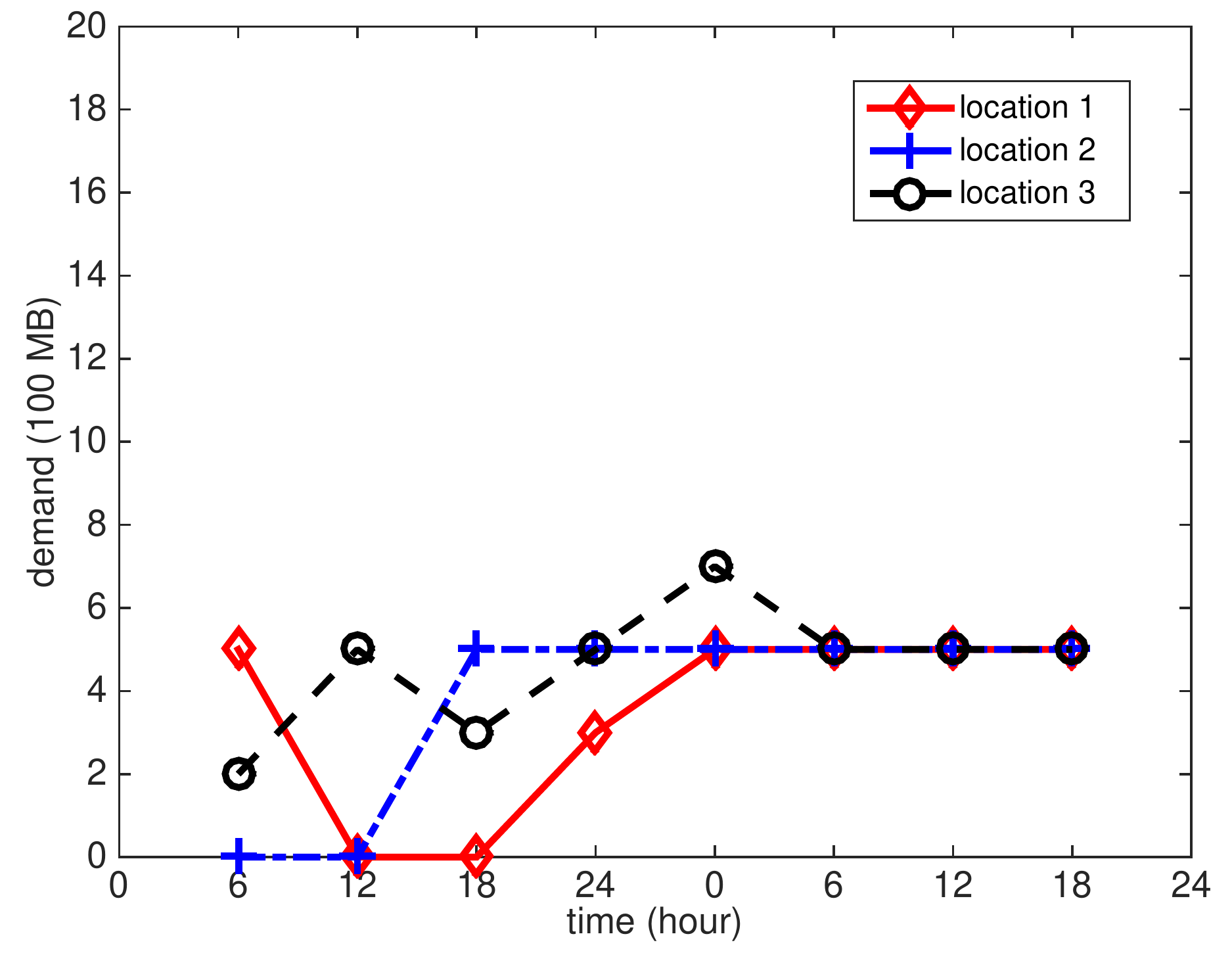}
      \caption{Usage under time and location aware pricing (the linear utility case)}\label{fig:NewDataTLPBCD}
    \end{minipage}
  \begin{minipage}[t]{0.25\linewidth}
  \centering
  \includegraphics[width=1\textwidth]{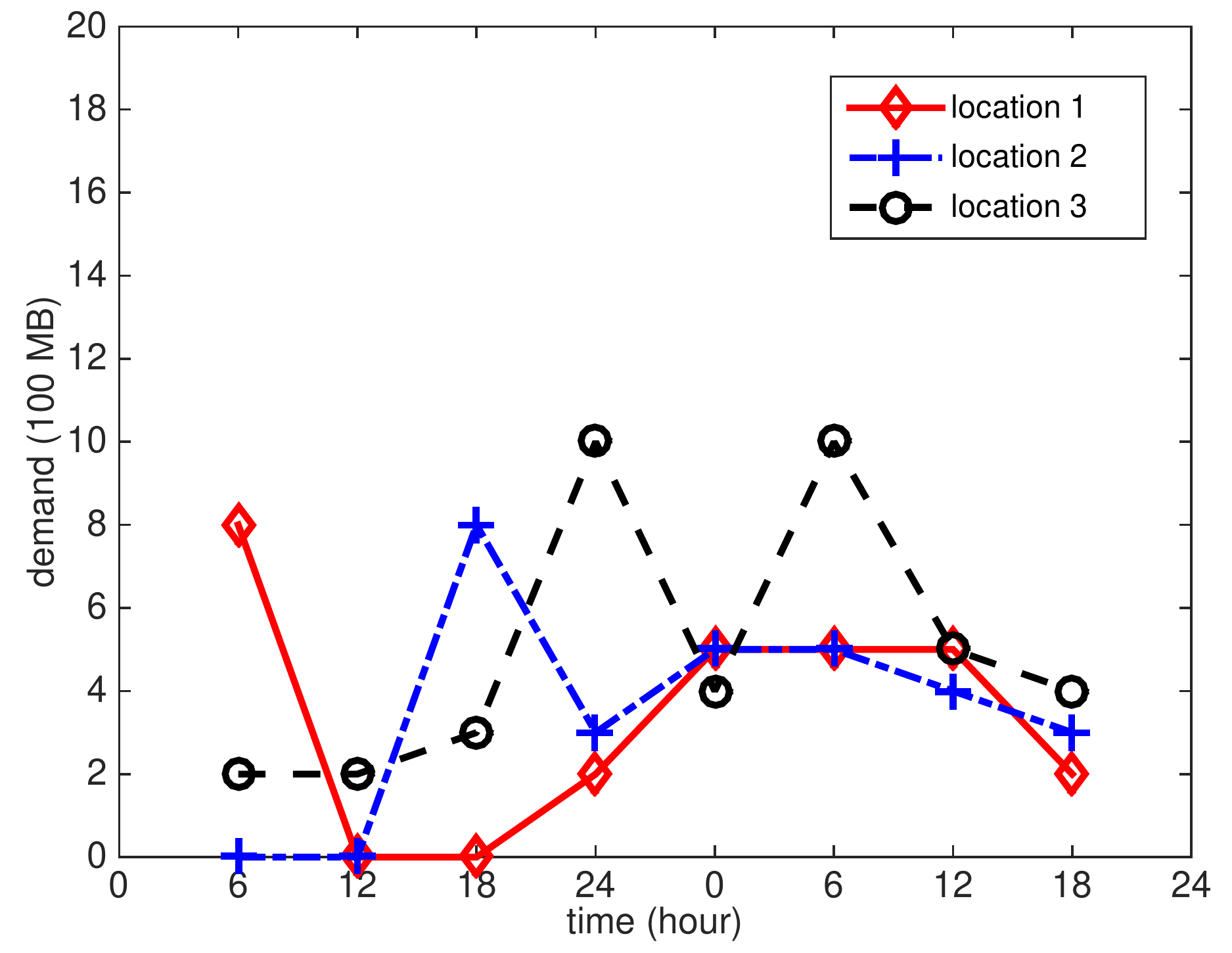}
  \caption{Usage under time-dependent pricing (the linear utility case)}\label{fig:NewDataTDPBCD}
  \end{minipage}
\end{figure*}

On one hand, by substituting $\overline{\lambda}_a(t,l)=k_a$ in (17) and (18), it is simple to check
$$x_a(t,l|t,l)+ \sum_{t'=t+1}^{T_t} \sum_{l'=1}^L \beta_a(t',l'|t,l) x_a(t',l'|t,l)=0,$$
which shows that $\overline{\lambda}_a(t,l)$ is an upper bound of the desired $\lambda_a^{\ast} (t,l).$ On the other hand, by substituting $$\underline{\lambda}_a(t,l)=\frac{k_a}{x_a^{ini}(t,l)+1}-p(t,l)$$ in (17) and (18), we obtain
\begin{align}
& x_a(t,l|t,l)+ \sum_{t'=t+1}^{T_t} \sum_{l'=1}^L \beta_a(t',l'|t,l) x_a(t',l'|t,l) \notag \\
\geq~ & x_a(t,l|t,l)=x_a^{ini}(t,l). \notag
\end{align}
This implies that $\underline{\lambda}_a(t,l)$ is a lower bound of the desired $\lambda_a^{\ast}(t,l).$

\section{Performance Comparison between the Time and Location Aware Pricing Scheme and the Time-Dependent Pricing Scheme}


Here we simulate a time-dependent pricing algorithm \cite{TUBE} and compare it with our proposed time and location aware pricing scheme. 

To see the comparisons clearly, we perform simulations on a simple example with $8$ time slots and $3$ locations. 
From Figure 13 in \cite{Traffic}, we can directly read the user data usage at three different base stations in one day. 
We aggregate the data every $3$ hours and get the following data usage pattern:
\begin{equation*}
\boldsymbol{x}_0 =
\left(
\begin{array}{cccccccc}
 8 & 0 & 0 & 2 & 8 & 3 & 5 & 2 \\
 0 & 0 & 8 & 3 & 16 & 2 & 4 & 3 \\
 2 & 2 & 3 & 10 & 4 & 1 & 7 & 2
\end{array}
\right).
\end{equation*} 
Matrix $\boldsymbol{x}_0$ represents the initial data usage in $8$ time slots at three locations under flat rate pricing, as plotted in Figure \ref{fig:InitialDataTDP}. 
The length of each time slot is $3$ hours.  
Each location corresponds to the coverage area of one base station. 
The unit of the data is $10^8$ bytes.

We construct the mobility profile based on \cite{M2} and \cite{MobilityData} as follows:
\begin{equation*}
\boldsymbol{\alpha} =
\left(
\begin{array}{cccccccc}
 0.2 & 0.15 & 0.1 & 0.3 & 0.4 & 0.4 & 0.3 & 0.3 \\
 0.1 & 0.05 & 0.2 & 0.4 & 0.5 & 0.4 & 0.3 & 0.2 \\
 0.7 & 0.8 & 0.7 & 0.3 & 0.1 & 0.2 & 0.4 & 0.5
\end{array}
\right).
\end{equation*} 

We further assume that the user's traffic scheduling interval $T=4$ and the network capacity $C=5$.

We first compare the performances of the two pricing schemes in the homogeneous logarithmic utility case. 
We assume that the cost parameter $\gamma=30$ and the delay tolerance parameter $\delta=0.6$.
Figure \ref{fig:NewDataTLPSPG} shows the new data usage pattern under the proposed time and location aware pricing scheme, and Figure \ref{fig:NewDataTDPSPG} shows the new data usage pattern under the time-dependent (but not location aware) pricing scheme. 
We can see that the data usage pattern under the time and location aware pricing scheme is smoother than the pattern under the time-dependent pricing scheme. 
Furthermore, the operator's cost reduction under the time and location aware pricing scheme is $97.58\%$, and the cost reduction under the time-dependent pricing scheme is $60.82\%$. 

We then compare the performances of the two pricing schemes in the homogeneous linear utility case. 
We assume that the cost parameter $\gamma=30$ and the delay tolerance parameter $\delta=0.95$. 
Figure \ref{fig:NewDataTLPBCD} shows the new data usage pattern under the proposed time and location aware pricing scheme, and Figure \ref{fig:NewDataTDPBCD} shows the new data usage pattern under the time-dependent (but not location aware) pricing scheme. 
Similarly, the data usage pattern under the time and location aware pricing scheme is smoother than the pattern under the time-dependent pricing scheme. 
Furthermore, the operator's cost reduction under the time and location aware pricing scheme is $97.06\%$, and the cost reduction under the time-dependent pricing scheme is $63.56\%$. 

\begin{IEEEbiography}[{\includegraphics[width=1in,height=1.25in,clip,keepaspectratio]{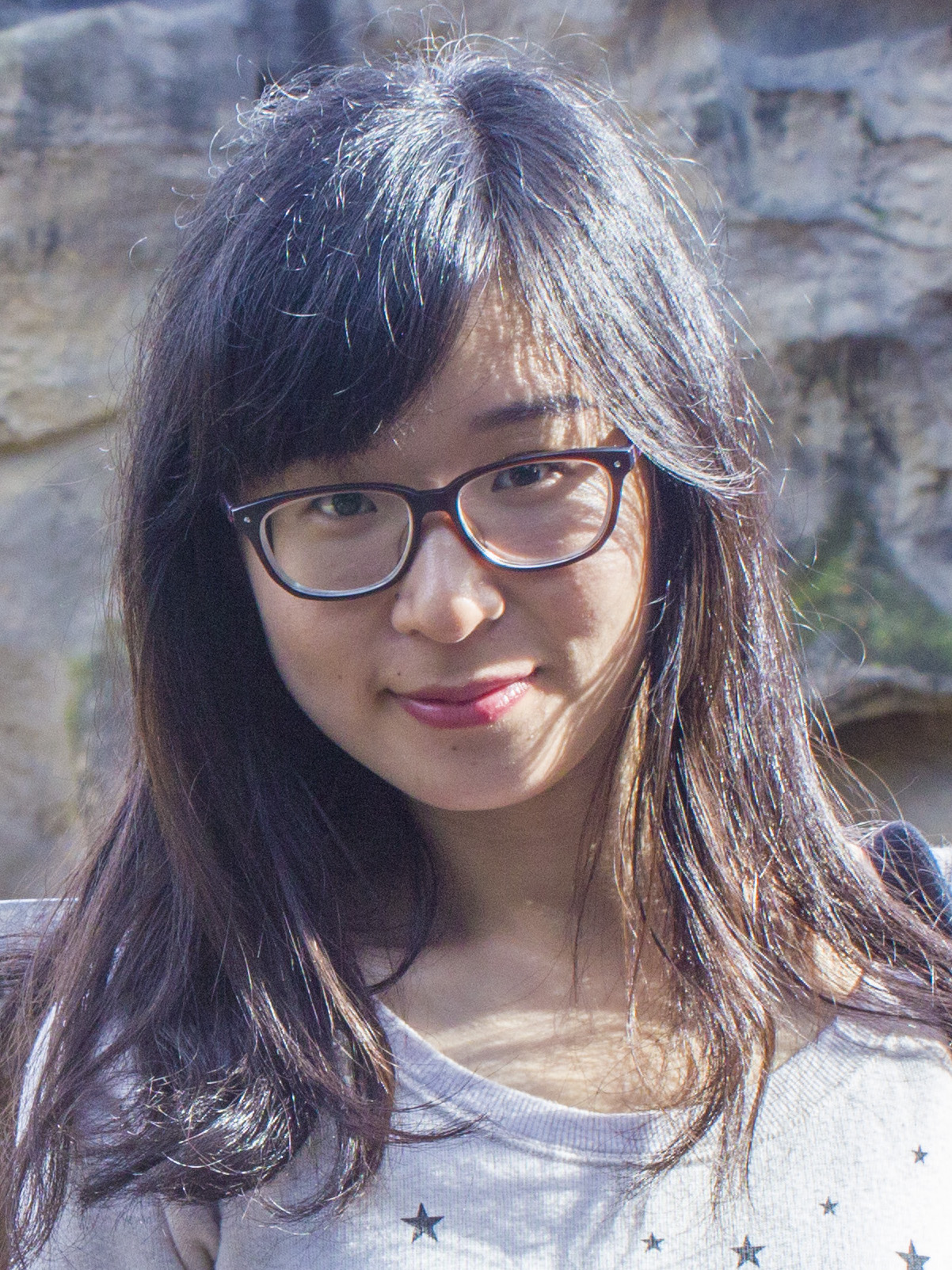}}]{Qian Ma}
(S'13) is a Ph.D. student in the Department of Information Engineering at the Chinese University of Hong Kong. 
She received the B.S. degree from Beijing University of Posts and Telecommunications (China) in 2012. 
Her research interests lie in the field of wireless communications and network economics. 
She is the recipient of the Best Student Paper Award from the IEEE International Symposium on Modeling and Optimization in Mobile, Ad
Hoc and Wireless Networks (WiOpt) in 2015.
\end{IEEEbiography}

\begin{IEEEbiography}[{\includegraphics[width=1in,height=1.25in,clip,keepaspectratio]{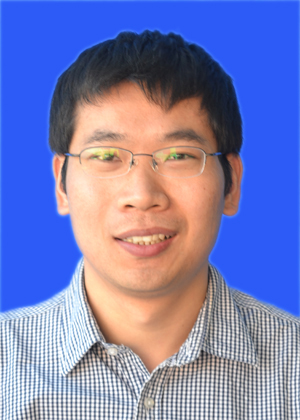}}]{Ya-Feng Liu}
(M'12) is an Assistant Professor of the Academy of Mathematics and Systems Science, Chinese Academy of Sciences, where he received the Ph.D degree in Computational Mathematics in 2012. 
His main research interests are nonlinear optimization and its applications to signal processing, wireless communications, and machine learning. 
He is a recipient of the Best Paper Award from the IEEE International Conference on Communications (ICC) in 2011 and the Best Student Paper Award from the International Symposium on Modeling and Optimization in Mobile, Ad Hoc and Wireless Networks (WiOpt) in 2015.
\end{IEEEbiography}

\begin{IEEEbiography}[{\includegraphics[width=1in,height=1.25in,clip,keepaspectratio]{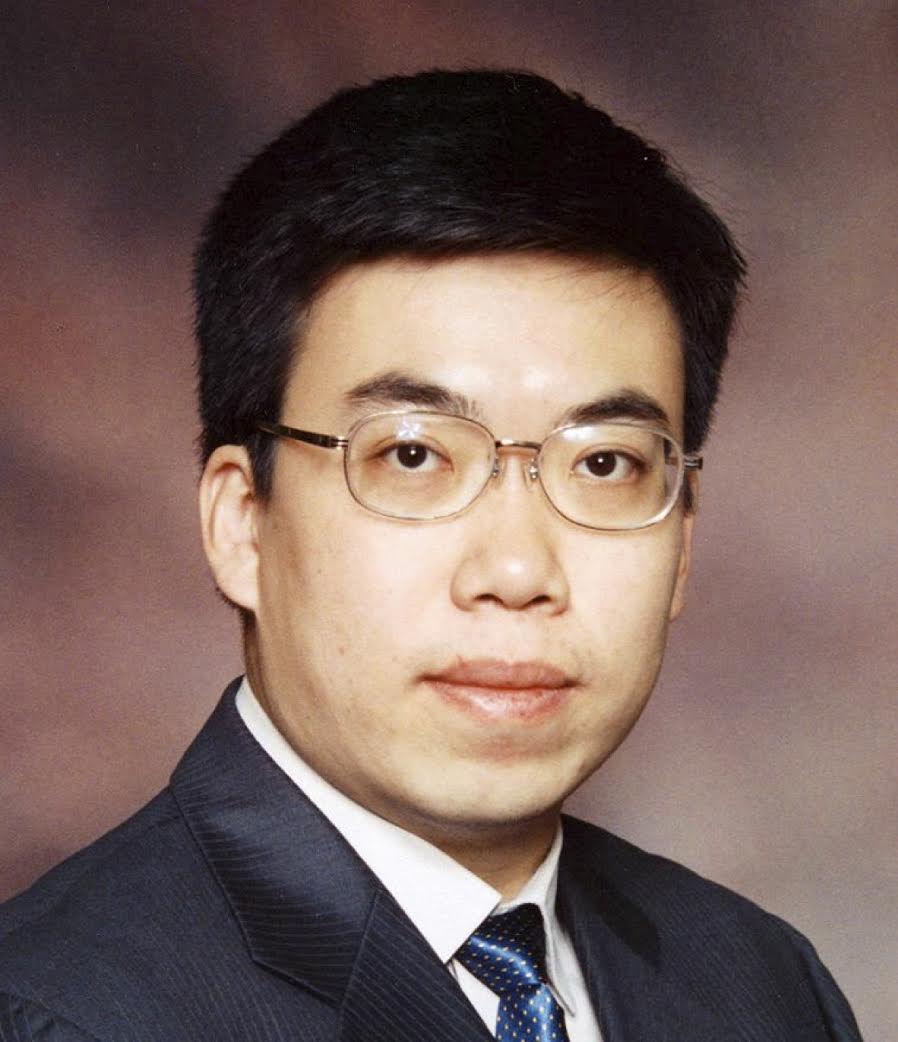}}]{Jianwei Huang}
(S'01-M'06-SM'11) is an Associate Professor and Director of the Network Communications and Economics Lab (ncel.ie.cuhk.edu.hk), in the Department of Information Engineering at the Chinese University of Hong Kong. 
He received the Ph.D. degree from Northwestern University in 2005. 
He is the co-recipient of 8 international Best Paper Awards, including IEEE Marconi Prize Paper Award in Wireless Communications in 2011. 
He has co-authored four books: ``Wireless Network Pricing,'' ``Monotonic Optimization in Communication and Networking Systems,''  ``Cognitive Mobile Virtual Network Operator Games,'' and ``Social Cognitive Radio Networks''. 
He has served as an Editor of several top IEEE Communications journals, including JSAC, TWC, and TCCN. 
He is a Senior Member and a Distinguished Lecturer of IEEE Communications Society.
\end{IEEEbiography}






\end{document}